\documentclass[preprint2]{aastex}          
\usepackage{graphicx}
\usepackage{psfrag}

\begin{document}

\title{Particle motion in the field of a five-dimensional charged black hole}

\shorttitle{Particle motion in the field of a five-dimensional charged black hole}
\shortauthors{S. Guha et al.}

\author{Sarbari Guha}
\affil{Department of Physics, St. Xavier's College (Autonomous), Kolkata 700 016}
\and
\author{Pinaki Bhattacharya\altaffilmark{1}}
\affil{Gopal Nagar High School, Singur 712409, West Bengal, India}
\and
\author{Subenoy Chakraborty}
\affil{Department of Mathematics, Jadavpur University, Kolkata 700032, India}


\altaffiltext{1}{Department of Physics, Jadavpur University, Kolkata 700032, India}

\begin{abstract}
In this paper, we have investigated the geodesics of neutral particles near a five-dimensional charged  black hole using a comparative approach. The effective potential method is used to determine the location of the horizons and to study radial and circular trajectories. This also helps us to analyze the stability of radial and circular orbits. The radius of the innermost stable circular orbits have also been determined. Contrary to the case of massive particles for which, the circular orbits may have up to eight possible values of specific radius, we find that the photons will only have two distinct values for the specific radii of circular trajectories. Finally we have used the dynamical systems analysis to determine the critical points and the nature of the trajectories for the timelike and null geodesics.
\end{abstract}


%

\section{Introduction}
\label{s:intro}

With the development of string theory, the study of black holes in higher-dimensional spacetimes have assumed fundamental importance \citep [see e.g,][]{ER, Kanti}. In these models, the minimal mass of a black hole can be much smaller than the Planck mass ($10^{19}$ GeV). Such mini black holes could be produced in high energy experiments: in colliders like LHC and in cosmic ray experiments \citep{collbh}. Estimations show that a TeV-size black hole in TeV gravity is small enough to be described within the classical solutions of vacuum Einstein equations, by an asymptotically flat black hole in higher dimensions, including the case of the rotating black holes. The study of higher dimensional black holes have gained momentum in the first decade of this millennium \citep{Gibbons,Sen,Frolov}. Static, spherically symmetric exterior vacuum solutions of the braneworld models were first proposed by Dadhich and others \citep{Dadhich,lrr}, with the metric having the structure of the Reissner-Nordstr\"{o}m (RN) solution and interpreted as a black hole with a tidal charge arising via gravitational effects from the fifth dimension. Static, spherically symmetric charged black hole solutions in 5-dimensions are known to be parameterized by their mass and electric charges and represent the RN solution \citep{ER}.

Extensive studies in higher-dimensional spacetimes over the last few decades have led many authors to investigate the geodesic motions in such spacetimes \citep{Page,Cardoso,Konoplya}. The geodesic structure of Schwarzschild AdS black hole has been considered by Cruz and others \citep{Cruz}. Motion of massive particles around a rotating black hole in a braneworld has been studied \citep{Abd} and the principal null geodesics and circular photon orbits for the Sen black hole have been investigated \citep{Hioki}. The effective potentials for radial null geodesics in RN-dS and Kerr-dS spacetimes were analyzed \citep{Stuch} and bound orbits in a Kerr spacetime have been discussed by some authors \citep{Kraniotis1}. Compact calculation of the perihelion precession of Mercury has been done, based on the inversion problem of hyperelliptic integrals by Kraniotis \citep{Kraniotis2}. Analytic solutions of the geodesic equations in various spacetime geometries in 4-dimensions, as well as a number of cases in higher dimensions has been obtained \citep{Hackmann1,Hackmann1a,Hackmann2,Kagra}. In this work, we have investigated the radial and circular trajectories for photons and massive particles in a five-dimensional RN spacetime using a comparative approach and have determined the fixed points of the phase trajectories. For such a non-rotating charged black hole, the solutions are uniquely characterized by their mass, charge and the cosmological constant \citep{Gibbons2}.

The paper is organized as follows: In Section II the line element and the horizon function has been defined and the horizons are located. The equations for the 5-dimensional geodesics in the region surrounding the black hole are determined in the next section. In Section IV, the trajectories of the test particles are analyzed using the method of effective Newtonian potential approximation. This equation is used to study radial motion and the corresponding stability of radial trajectories. To study circular motion of particles and the stability of circular orbits, the equations have been transformed in terms of the inverse of the radial coordinate and the energy and angular momentum of massive particles have been determined. The radius of the innermost stable circular orbit of massive particles and photons have been calculated. Contrary to the case of massive particles, for which the orbits may have up to eight possible values of the specific radius, it is found that photons can have circular trajectories for only two distinct values of specific radius of the orbits. In Section V, we have made an analysis of the timelike and null geodesics using the dynamical systems approach. The fixed points are determined and the nature of the phase trajectories have been analyzed. The summary and the conclusions are presented in Section VI.

\section{Preliminaries}
\label{s:1}

Let us consider a 5-dimensional spacetime in the presence of a cosmological constant. The field equations are given by
\begin{equation}\label{01}
\bar{G}_{AB} = - \Lambda_{(5)}\bar{g}_{AB} + \kappa^{2}_{(5)}\bar{T}_{AB}
\end{equation}
where $\bar{g}_{AB}$ is a 5-dimensional metric of signature (- + + + +), $\bar{G}_{AB}$ is the 5-dimensional Einstein tensor, $\bar{T}_{AB}$ represents the 5-dimensional energy-momentum tensor and $\Lambda_{(5)}$ is the 5-dimensional cosmological constant. The spacetime is assumed to be one of constant curvature $\bar{K}=\frac{\beta}{l^2}$. For AdS geometry, $K$ is negative and $\beta=-1$, whereas for dS, $K$ is positive so that $\beta=1$. The radius of curvature
\begin{equation}\label{03}
l= \sqrt{ \frac{3\beta}{\Lambda_{(5)}}}
\end{equation}
of the spacetime provides the length scale necessary to have a horizon. To simplify our notation we shall drop the subscript and henceforth denote $\Lambda_{(5)}$ by $\Lambda$. The above equations are satisfied by the exterior metric of the black hole field, which has the form
\begin{equation}\label{04}
dS^{2}= -f(r)dt^2 + \frac{dr^2}{f(r)} + r^2 d\Omega_{3}^2,
\end{equation}
where, $d\Omega_{3}^2=(d\theta^2 + sin^2 \theta(d\phi^2 + sin^2 \phi d\psi^2))$ is the metric of the unit 3-sphere. For this static, spherically symmetric exterior vacuum solution of the Einstein equations in RN spacetime, the lapse function f(r) is defined as
\begin{equation}\label{05}
f(r) = 1 - \left(\frac{2M}{r}\right)^2 + \left( \frac{q^2}{r^2} \right)^2 - \frac{\Lambda r^2}{6} = \frac{\triangle}{r^4}.
\end{equation}
Here, $q$ and $M$ are the charge and the mass of the black hole, respectively. For a given value of $M$, $q$ and $\Lambda$, the \emph{horizon function} $\triangle$ depends only on the radial coordinate $r$. In diagonal form, the exterior metric can be written as
\begin{equation}\label{05a}
dS^{2}= -\frac{\triangle}{r^4}dt^2 + \frac{r^4}{\triangle}dr^2 + r^2 d\Omega_{3}^2.
\end{equation}
The spacetime, has an intrinsic singularity at $r=0$, the nature of which depends on the choice of $\Lambda$ and $q$. These singularities may either be black holes or naked singularities. In fact there is no guarantee that "black hole" candidates are indeed black holes and some (or all) of "black hole" candidates could instead be naked singularities \citep{Virbhadra1,Virbhadra2}. Further, the net electric charge of astronomical objects are constrained within a certain range if their exterior spacetimes are described by the Reissner-Nordstr\"{o}m metric, which induces power-law potentials \citep{Iorio}. Here we assume that both $\Lambda$ and $q$ are chosen in such a way that the spacetime do not have any spacelike naked singularity \citep{Carroll}. The lapse function vanishes at the zeros of the equation
\begin{equation}\label{06}
\triangle = 0,
\end{equation}
for which
\begin{equation}\label{07}
\Lambda r^6 -6r^4 + 24M^2r^2 -6q^4 = 0.
\end{equation}

\begin{figure}[tb]
\psfrag{V_eff}{\scalebox{1}{$V_eff$}}
\psfrag{h}{\scalebox{1}{$r/M$}}
\includegraphics[scale=.30]{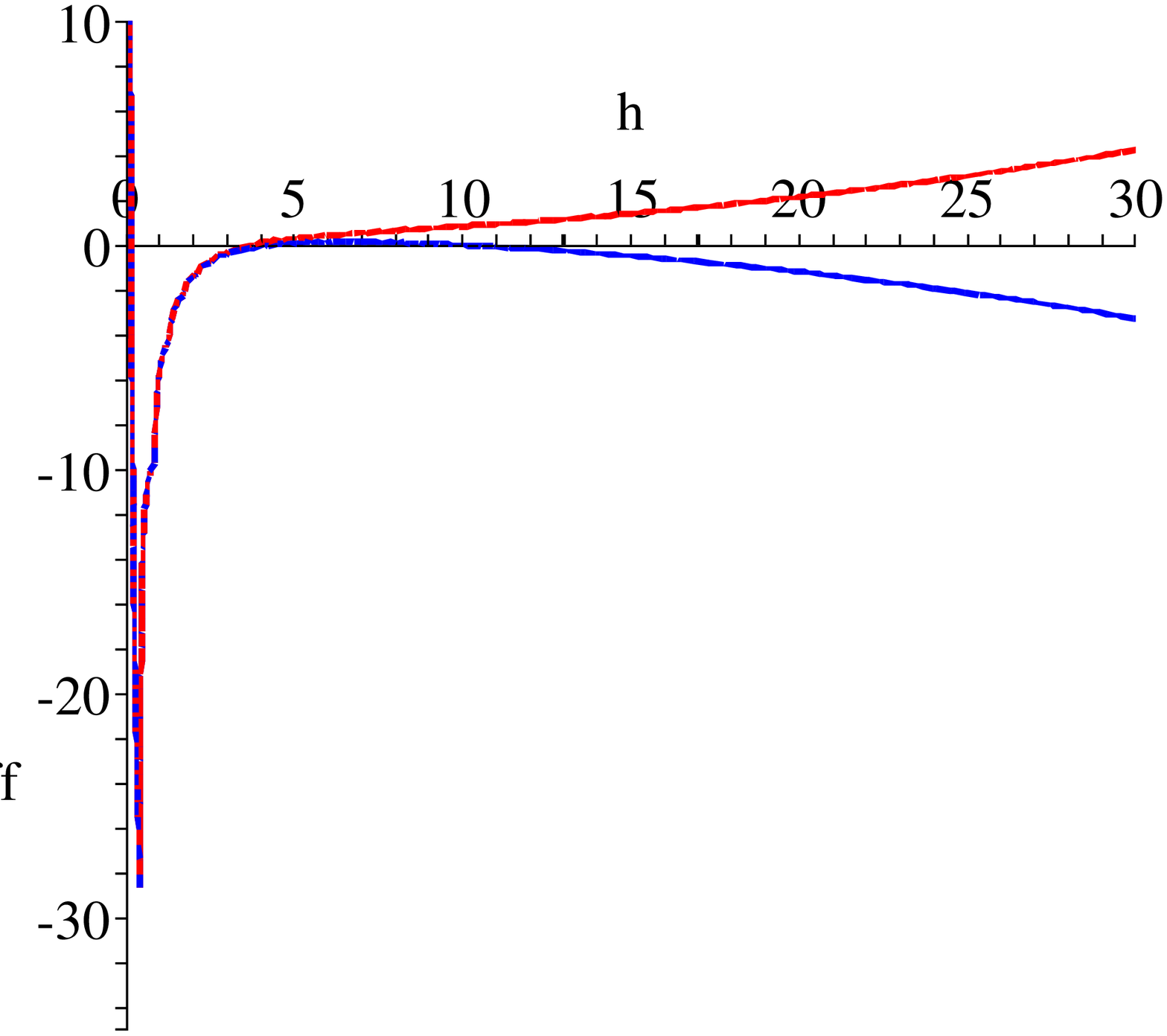}
\includegraphics[scale=.30]{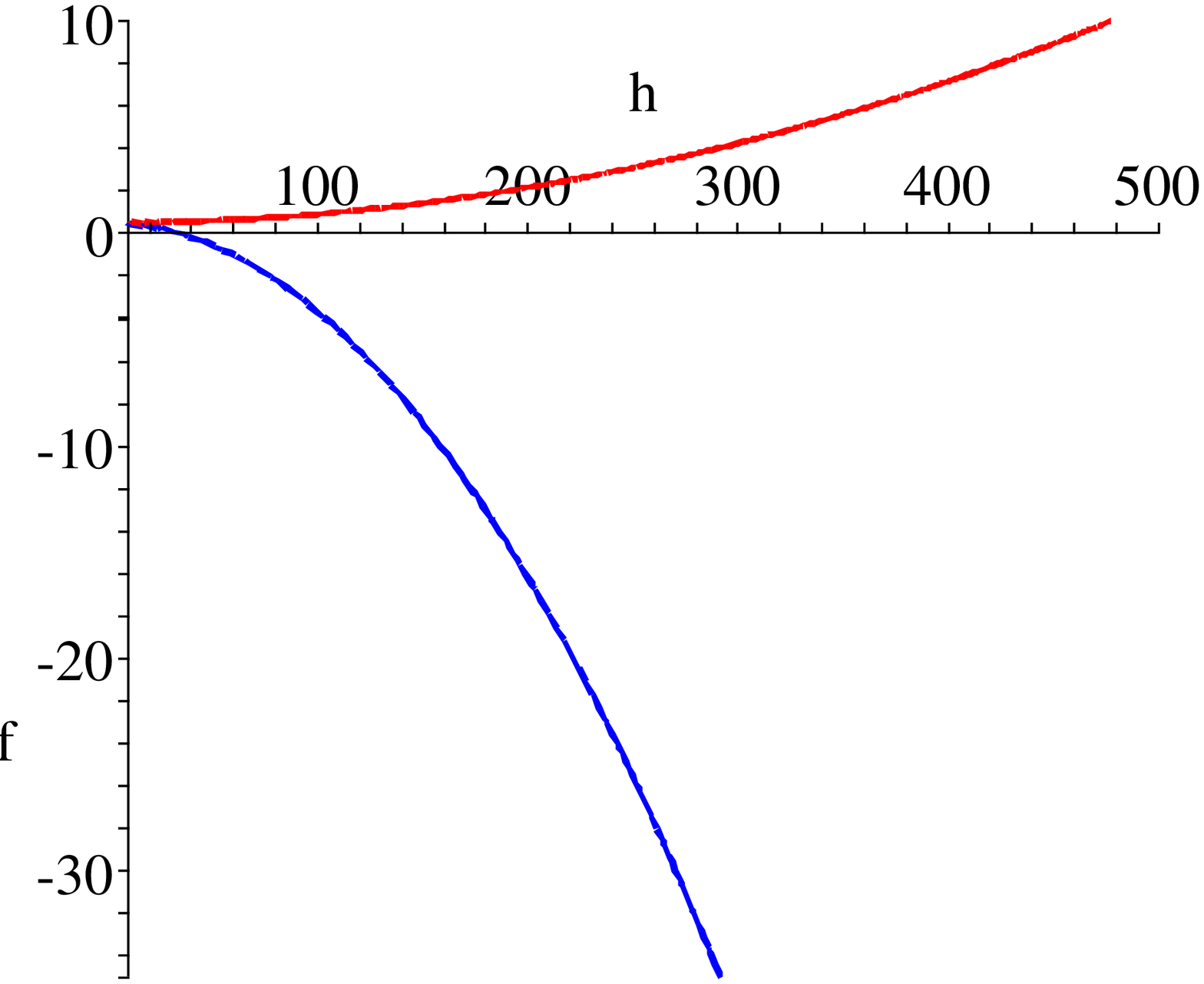}
\caption{Diagram showing the plot of effective potential $V_{eff}$ vs radial distance $r/M$ in the case of radial motion of massive particles with $q/M =0.5$ for different range of values of $r/M$ for sample values of $\Lambda=-0.05$ (red) and $\Lambda=0.05$ (blue) in the top figure and $\Lambda=-0.0005$ (red) and $\Lambda=0.005$ (blue) in the bottom figure}
\label{effpot0}
\end{figure}

The horizons are located at the real, positive zeros of $\triangle$, indicating the coordinate singularities. The effective potential also vanishes at the zeros of the horizon function $\triangle$, indicating the location of these horizons. The variation of the effective potential $V_{eff}$ with specific radius $r/M$ in the case of radial motion of massive particles, in the field of the black hole for different range of values of $r/M$, are shown in Fig.~\ref{effpot0}. The location of the horizons can be easily identified from these figures. As expected, we find that the field of a charged black hole in the de Sitter spacetime is characterized by the presence of all three horizons: the Cauchy horizon, the event horizon and the cosmological horizon, whereas there are only two horizons in the Anti de Sitter spacetime.

Among the six distinct roots of equation (\ref{06}) in the RN spacetime, only two are real for the AdS case, with a single one being positive, which gives us the radius of the event horizon of the black hole. This positive root is found as
\begin{equation}\label{07a}
r = \sqrt{ \left( \frac{2}{\Lambda} - \frac{(242^{1/3}M^2)}{\Re} + \frac{122^{1/3}}{\Lambda \Re} + \frac{\Re}{32^{1/3} \Lambda}   \right)  }
\end{equation}

\begin{flushleft}
where
$\Re = 3 \left( 2A + 2\sqrt{ B^3 + A^2 }\right)^{1/3}$

with $A = 8 - 24 M^2 \Lambda + 3 q^4 \Lambda^2$ and $B = 8 M^2 \Lambda - 4$.
\end{flushleft}

\section{Five-dimensional Geodesics}
\label{s:2}

The geodesic equations are
\begin{equation}\label{08}
\frac{d^{2}z^{A}}{d\lambda^{2}}+^{(5)}\Gamma^{A}_{BC}\frac{dz^{B}}{d\lambda}\frac{dz^{C}}{d\lambda}=0,
\end{equation}
where $\lambda$ is an affine parameter along the geodesic curve $z^{A}(\lambda)$, $ ^{(5)}\Gamma^{A}_{BC}$ are the 5-dimensional Christoffel symbols of the second kind and ${z^{A}}$ are the coordinates of the 5-dimensional spacetime. The nature of trajectories of the test particles depend on their energies and angular momentum as well as the cosmological constant and the electric charge of the black hole. The mass of the black hole can be absorbed by rescaling the radial coordinate. We consider only neutral test particles, so that their trajectories describe geodesics in the RN spacetime \citep{Chandra}.

On account of the spherical symmetry, the motion of the test particles can be determined by analyzing their motion on the equatorial plane, for which $\theta,\phi=\pi/2$. Substituting the explicit form of the bulk metric, we obtain the following set of geodesic equations

\begin{equation}\label{e6}
\frac{d^{2} t}{d \lambda^{2}} + \frac{B(r)}{A(r)}\frac{dt}{d\lambda}\frac{dr}{d\lambda} = 0,
\end{equation}
\begin{eqnarray*}
\frac{d^{2} r}{d \lambda^{2}}+ A(r)B(r)\left( \frac{dt}{d\lambda} \right)^2 - \frac{B(r)}{A(r)}\left( \frac{dr}{d\lambda} \right)^2 + rA(r) \left( \frac{d\psi}{d\lambda}\right)^{2}
\end{eqnarray*}
\begin{equation}\label{e7}
\qquad \qquad \qquad \qquad \qquad \qquad \qquad = 0,
\end{equation}
\begin{equation}\label{e8}
\frac{d^{2}\psi}{d\lambda^{2}} + \frac{1}{r}\frac{dr}{d\lambda}\frac{d\psi}{d\lambda} = 0.
\end{equation}
where we have defined $A(r)=-f(r)$ and $B(r)= \frac{1}{r}\left[- \left(\frac{2M}{r}\right)^2 + 2\left( \frac{q^2}{r^2} \right)^2 + \frac{\Lambda r^2}{6} \right]$.

We now proceed to make a study of the geodesics using a comparative approach. The effective Newtonian potential formalism is used to determine the location of the horizons and to study radial and circular trajectories. Most of the papers mentioned earlier from \citep{Cruz} to \citep{Hackmann2} use this method. Subsequently, we determine the critical points and the nature of the trajectories of the particles using the dynamical systems analysis \citep{dynsys1,dynsys2}.

\section{Analysis of particle trajectories using the effective Newtonian potential approach}
\label{s:3}

The lagrangian for particle motion in this RN-AdS$_{5}$ is given by

\begin{eqnarray*}
\textsf{{\L}} = g_{AB}\dot{z}^{A}\dot{z}^{B} = - \left[ 1 - \left(\frac{2M}{r}\right)^2 + \left( \frac{q^2}{r^2} \right)^2 - \frac{\Lambda r^2}{6} \right]\dot{t}^2
\end{eqnarray*}
\begin{eqnarray*}
+ \frac{1}{1 - \left(\frac{2M}{r}\right)^2 + \left( \frac{q^2}{r^2} \right)^2 - \frac{\Lambda r^2}{6}} \dot{r}^2
\end{eqnarray*}
\begin{equation}\label{09}
+ r^2( \dot{\theta}^2 + sin^2 \theta \dot{\phi}^2 + sin^2 \theta sin^2 \phi \dot{\psi}^2),
\end{equation}
where, an overdot represents differentiation with respect to the affine parameter $\lambda$ along the geodesic $z^{A}(\lambda)$. Since $t$ and $\psi$ are cyclic coordinates, there are two conserved quantities, namely the energy $E$ and the momentum conjugate to $\psi$. Thus, we have
\begin{equation}\label{11}
E = g_{tt}\frac{dt}{d \lambda} = -f(r)\frac{dt}{d \lambda} = A(r)\frac{dt}{d \lambda},
\end{equation}
and
\begin{equation}\label{10}
p_{\psi}=\frac{\partial \textsf{{\L}}}{\partial\dot{\psi}}= 2r^2\dot{\psi} =\verb"constant" = 2L.
\end{equation}
so that
\begin{equation}\label{12}
L = r^2\frac{d \psi}{d \lambda},
\end{equation}
where $L$ is now the total angular momentum of the particles confined to the equatorial plane. From the normalization condition
\begin{equation}\label{13}
g_{AB}\frac{dz^{A}}{d\lambda}\frac{dz^{B}}{d\lambda} = - \epsilon,
\end{equation}
where, $\epsilon=1$ for timelike geodesics and $\epsilon=0$ for null geodesics, we have for the geodesics on the equatorial plane
\begin{equation}\label{14}
\left (\frac{dr}{d \lambda} \right)^2 = E^{2} - \frac{\triangle}{r^4} \left(\epsilon + \frac{L^2}{r^2} \right) = E^{2} + A(r)\left(\epsilon + \frac{L^2}{r^2} \right).
\end{equation}
Here we have used the sign convention of Carroll \citep{Carroll} and Hartle \citep{Hartle}. We can rewrite (\ref{14}) as
\begin{equation}\label{14a}
\frac{1}{2}\left (\frac{dr}{d \lambda} \right)^2 = E_{eff} - V_{eff}(r),
\end{equation}
where the equatorial geodesics defined above are subjected to an effective potential $V_{eff}(r)$ given by
\begin{eqnarray*}
V_{eff}(r) = \frac{\triangle}{2r^4} \left(\epsilon + \frac{L^2}{r^2} \right)
\end{eqnarray*}
\begin{equation}\label{15}
= \frac{1}{2}\left( 1 - \left(\frac{2M}{r}\right)^2 + \left( \frac{q^2}{r^2} \right)^2 - \frac{\Lambda r^2}{6} \right) \left(\epsilon + \frac{L^2}{r^2} \right),
\end{equation}
and
\begin{equation}\label{16}
E_{eff} = \frac{1}{2}E^2.
\end{equation}
Thus (\ref{14a}) is the equation of motion of a particle of unit mass and effective energy $E_{eff}$, moving in a one-dimensional potential $V_{eff}(r)$. Since $r$ should be real and positive, the physically acceptable regions are given by those $r$ for which $E_{eff} > V_{eff}(r)$, owing to the square on the left hand side of (\ref{14a}). Substituting for $A(r)$ in (\ref{14}), we arrive at the equation
\begin{eqnarray*}
\left (\frac{dr}{d \lambda} \right)^2 = E^{2} - \epsilon + \frac{\Lambda \epsilon r^2}{6} + \frac{\Lambda L^2}{6} + \frac{4\epsilon M^2}{r^2} - \frac{L^2}{r^2}
\end{eqnarray*}
\begin{equation}\label{17}
 + \frac{4M^2 L^2}{r^4} - \frac{\epsilon q^4}{r^4} - \frac{L^2q^4}{r^6}.
\end{equation}
The shape of the trajectories can be determined by using (\ref{12}) to express $\dot{r}$ in the following form:
\begin{equation}\label{18}
\frac{dr}{d\lambda} = \frac{dr}{d\psi}\frac{d\psi}{d\lambda} = \frac{L }{r^2} \frac{dr}{d\psi}.
\end{equation}
The two special cases of particle trajectories, which are important from the physical point of view are the radial motion and the circular motion of the particles. Equation (\ref{14a}) is used to study radial free fall and the stability of radial trajectories. However, analysis of the circular motion of particles is more conveniently done if we introduce the variable change \citep{Hobson},
\begin{equation}\label{19}
u=r^{-1},
\end{equation}
so that
\begin{eqnarray*}
\left( \frac{du}{d\lambda} \right)^2 = -L^2 q^4 u^{10} + (4 M^2 L^2 -\epsilon q^4)u^8
\end{eqnarray*}
\begin{equation}\label{21}
+ (4 \epsilon M^2 - L^2)u^6 + \left(E^2 - \epsilon + \frac{\Lambda L^2}{6} \right)u^4 + \frac{\Lambda \epsilon u^2}{6}.
\end{equation}

Using equations (\ref{18}) to (\ref{21}), the equation defining the circular trajectories of test particles in the field of a non-rotating charged black hole in a five-dimensional spacetime, is obtained as

\begin{eqnarray*}
\left( \frac{du}{d\psi} \right)^2 = - q^4 u^6 + \left( 4 M^2 - \frac{\epsilon q^4}{L^2} \right)u^4 + \left( \frac{4 \epsilon M^2}{L^2} - 1 \right)u^2
\end{eqnarray*}
\begin{equation}\label{22}
 + \left( \frac{E^2 - \epsilon}{L^2} + \frac{\Lambda}{6} \right) + \frac{\Lambda \epsilon }{6 u^2 L^2} = P(u).
\end{equation}

\subsection{Radial motion}

For radial motion, $\psi=constant$ and hence $L=0$. Therefore we consider equation (\ref{14}) in the form
\begin{equation}\label{23}
\left (\frac{dr}{d \lambda} \right)^2 = E^{2} + A(r)\epsilon,
\end{equation}
to analyze the radial trajectories.

\subsubsection{Motion of massive particles}

In case of massive particles, (\ref{23}) gives
\begin{equation}\label{24}
\left (\frac{dr}{d \lambda} \right)^2 = E^{2} + A(r)= E^{2} - 1 + \left(\frac{2M}{r}\right)^2 - \left( \frac{q^2}{r^2} \right)^2 + \frac{\Lambda r^2}{6}.
\end{equation}
On differentiating (\ref{24}) with respect to $\lambda$ and dividing by $2\dot{r}$, we get
\begin{equation}\label{25}
\frac{d^{2} r}{d \lambda^{2}} = - \frac{4 M^2}{r^3} + \frac{2 q^4}{r^5} + \frac{\Lambda r}{6}.
\end{equation}

\begin{figure}[tb]
\psfrag{V_eff}{\scalebox{1}{$V_eff$}}
\psfrag{h}{\scalebox{1}{$r/M$}}
\includegraphics[scale=.30]{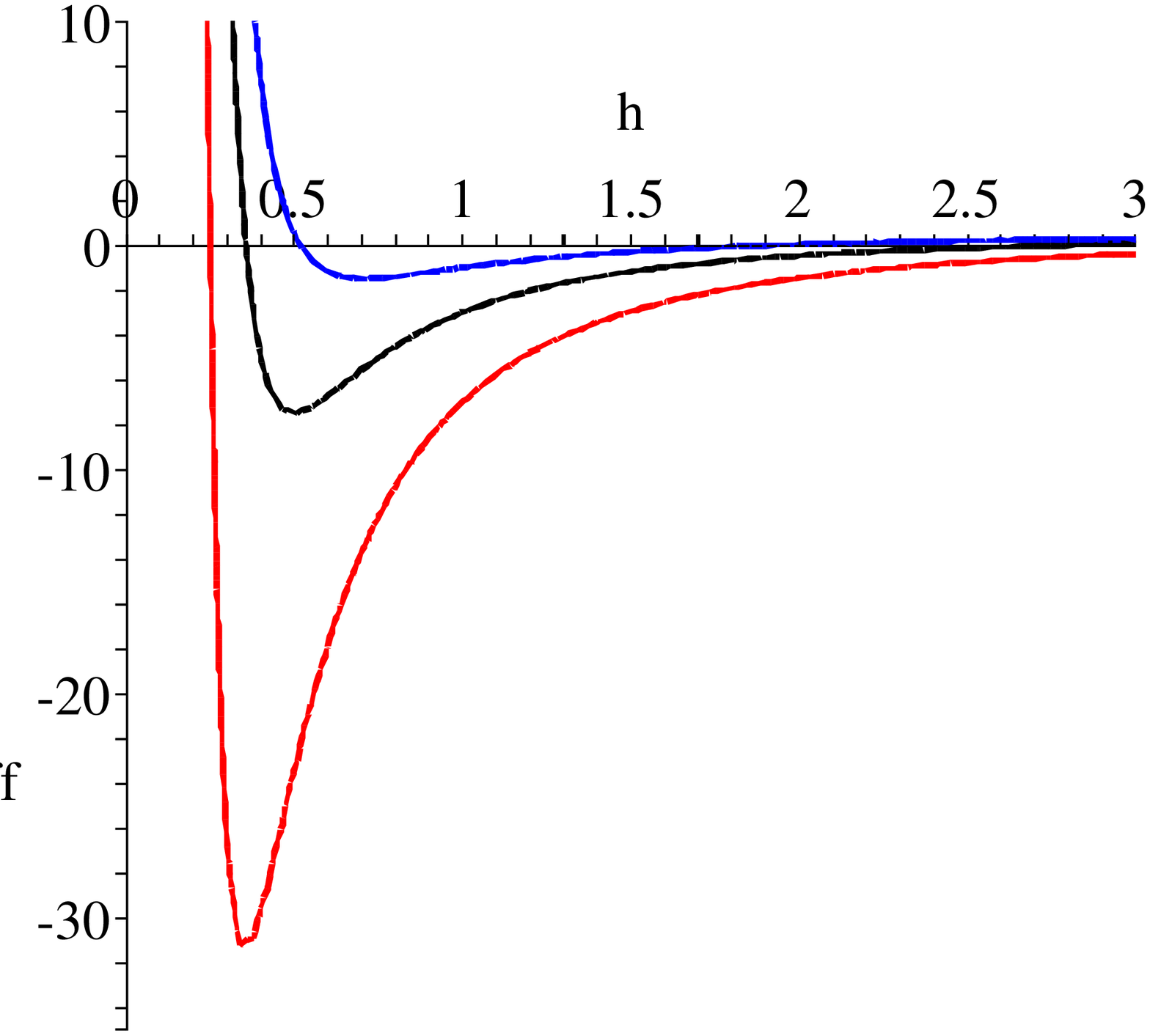}
\caption{Diagram showing the plot of effective potential $V_{eff}$ for $q/M =0.5$ (red), $q/M = 1/\sqrt{2}$ (black) and $q/M = 1$ (blue) in the case of radial motion of massive particles}
\label{effpot}
\end{figure}

Choosing the affine parameter $\lambda$ to be the proper time $\tau$ along the path, we find that the force per unit mass for massive particles will be attractive when
\begin{equation}\label{26}
\frac{4 M^2}{r^3} > \frac{2 q^4}{r^5} + \frac{\Lambda r}{6},
\end{equation}
which is the condition necessary for the existence of bound states. As the particle moves in the field of the black hole, it gains kinetic energy from the energy of gravitational interaction. The corresponding change in the gravitational potential energy of the particle can be estimated by considering that the particle starts from rest at a position where the radial coordinate is $R$. Writing equation (\ref{24}) in terms of the proper time $\tau$ of the particle, and assuming that at $r=R$, $dr/d\tau = 0$, we obtain the kinetic energy per unit mass gained by the particle in terms of the change in its gravitational potential energy, as follows:
\begin{eqnarray*}
\frac{1}{2}\left( \frac{dr}{d \tau} \right)^2 = 2M^2 \left(\frac{1}{r^2} - \frac{1}{R^2} \right) - \frac{q^4}{2}\left( \frac{1}{r^4} - \frac{1}{R^4} \right)
\end{eqnarray*}
\begin{equation}\label{28}
 + \frac{\Lambda}{12}(r^2 - R^2).
\end{equation}
Fig.~\ref{effpot} shows the plot of $V_{eff}$ vs the specific radius $r/M$ of the orbits, for different values of the specific charge $q/M$ of the black hole, where we have
\begin{eqnarray*}
V_{eff}(r) = \frac{1}{2}\left( 1 - \left(\frac{2M}{r}\right)^2 + \left( \frac{q^2}{r^2} \right)^2 - \frac{\Lambda r^2}{6} \right)
\end{eqnarray*}
for the radial motion of massive particles.

\begin{figure}[tb]
\psfrag{V_eff}{\scalebox{1}{$V_eff$}}
\psfrag{h}{\scalebox{1}{$r/M$}}
\includegraphics[scale=.30]{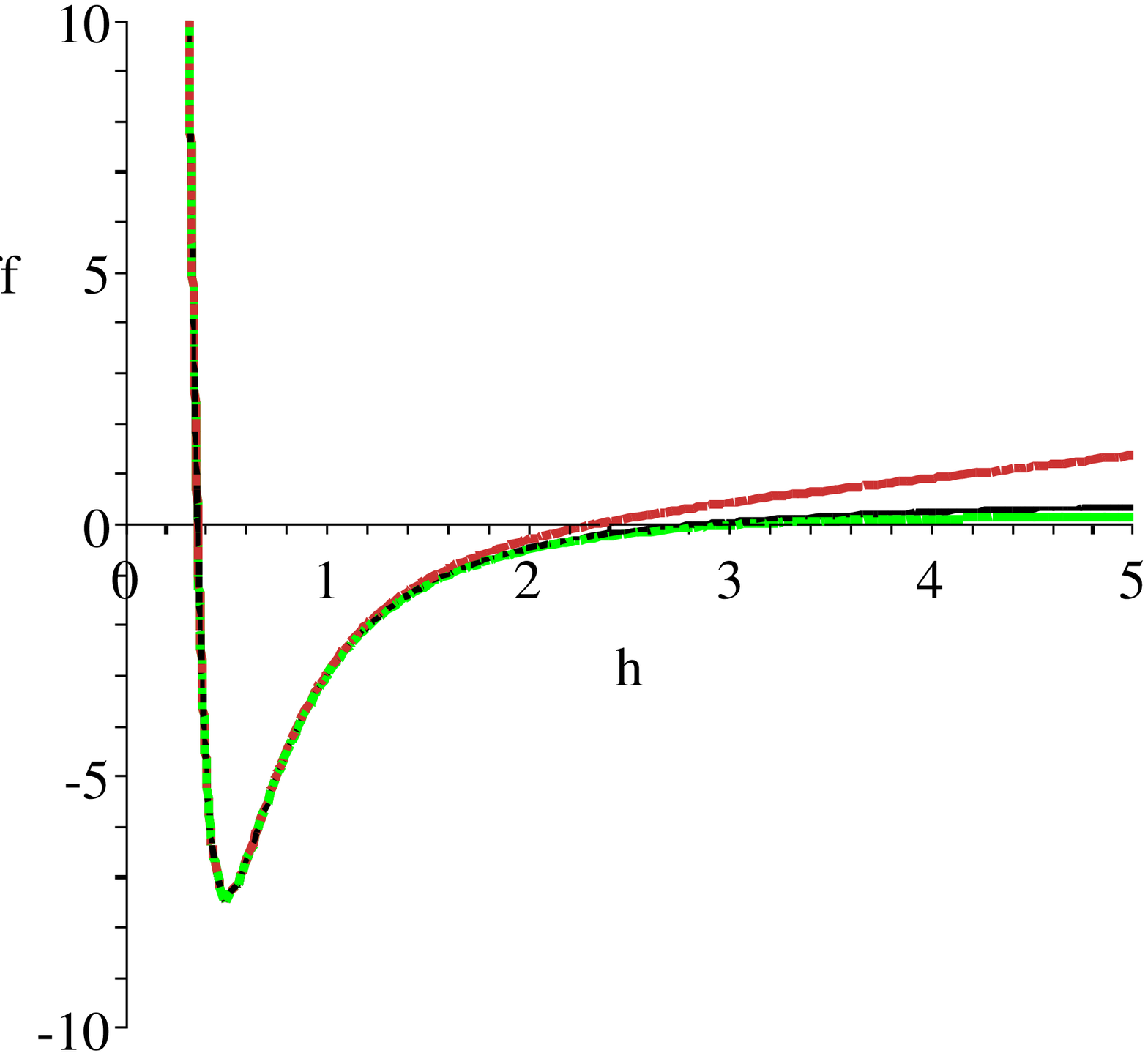}
\includegraphics[scale=.30]{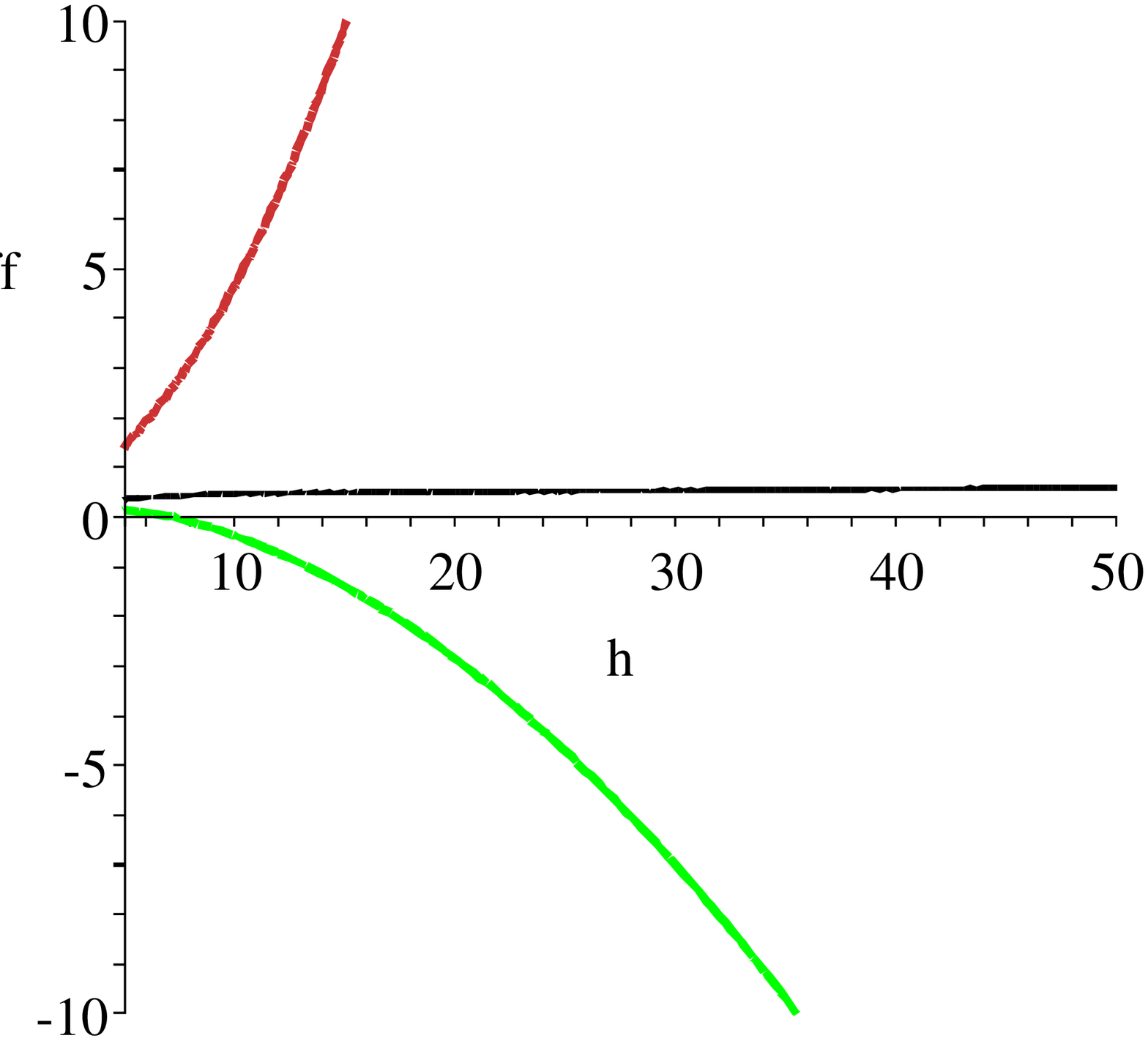}
\caption{Diagram showing the plot of effective potential $V_{eff}$ over different range of values of $r/M$ for sample values of $\Lambda=-0.5$ (orange) $\Lambda=-0.0005$ (black) and $\Lambda=0.1$ (green) in the case of radial motion of massive particles having $q/M = 1/\sqrt{2}$}
\label{effpot1}
\end{figure}

The variation of the effective potential for different values of $\Lambda$ with $q/M = 1/\sqrt{2}$ is shown in Fig.~\ref{effpot1}. The variation is similar to the Schwarzschild case, as illustrated in \citep{Hackmann1a}. The similarity is due to the fact that the Reissner–Nordstr\"{o}m metric differs from the Schwarzschild metric only in the definition of the \emph{horizon function} $\triangle$ (on account of the charge of the black hole) and the degree of the polynomial $P(u)$.

Using (\ref{11}) and (\ref{24}) we find that
\begin{eqnarray*}
\frac{dr}{dt} = \frac{1}{E}\left( 1 - \left(\frac{2M}{r}\right)^2 + \left( \frac{q^2}{r^2} \right)^2 - \frac{\Lambda r^2}{6} \right)
\end{eqnarray*}
\begin{equation}\label{29}
\times \left( E^2 - 1 + \left(\frac{2M}{r}\right)^2 - \left( \frac{q^2}{r^2} \right)^2 + \frac{\Lambda r^2}{6} \right)^{1/2}.
\end{equation}

Hence, at large distances from the black hole, the above derivative will blow up due to the contribution of the $\Lambda$-term. Thus the particles will not be able to reach there under the effect of the black hole field. However, at finite distance from the source, both the time derivative and the energy of the particles must be finite, with the exact trajectories on the $(t,r)$-plane being determined by this equation.

\subsubsection{Motion of photons}

The equation for the radial motion of photons is given by
\begin{equation}\label{33}
\left (\frac{dr}{d \lambda} \right)^2 = E^{2} ,
\end{equation}
where $E$ is defined by (\ref{11}). Writing
\begin{eqnarray*}
\frac{dr}{dt} &=& \frac{dr}{d \lambda} \frac{d \lambda}{dt}
\end{eqnarray*}
we have
\begin{equation}\label{34}
\left( \frac{dr}{dt} \right) =  \pm A(r) = \mp \left[1 - \left(\frac{2M}{r}\right)^2 + \left( \frac{q^2}{r^2} \right)^2 - \frac{\Lambda r^2}{6} \right].
\end{equation}
The trajectory of the particles can be represented on the $(t,r)$-plane after integrating the above equation. The solution involves inverse functions. However, to determine the critical points and the nature of the particle trajectories, we have to use the dynamical systems analysis. The trajectories will also be determined from this analysis, which will be taken up in the next section.

\subsection{Circular motion}

For equilibrium circular orbits, $u=constant$ in (\ref{22}) and hence $P(u)=0$, as well as $P^{\prime}(u)=0$. From (\ref{22}) we have
\begin{eqnarray*}
P(u) = - q^4 u^6 + \left( 4 M^2 - \frac{\epsilon q^4}{L^2} \right)u^4 + \left( \frac{4 \epsilon M^2}{L^2} - 1 \right)u^2
\end{eqnarray*}
\begin{equation}\label{35}
+ \left( \frac{E^2 - \epsilon}{L^2} + \frac{\Lambda}{6} \right) + \frac{\Lambda \epsilon }{6 u^2 L^2}.
\end{equation}
Thus
\begin{eqnarray*}
P^{\prime}(u) = - 6q^4 u^5 + 4\left( 4 M^2 - \frac{\epsilon q^4}{L^2} \right)u^3
\end{eqnarray*}
\begin{equation}\label{36}
+ 2\left( \frac{4 \epsilon M^2}{L^2} - 1 \right)u - \frac{2\Lambda \epsilon }{6 u^3 L^2}.
\end{equation}

From (\ref{36}), applying the condition $P^{\prime}(u)=0$, we obtain the expression for the angular momentum $L$ of a particle moving in a circular orbit of radius $r=1/u$ in RN-AdS$_{5}$ as follows:
\begin{equation}\label{37}
L^2 = \frac{\epsilon [ 6 q^4 u^6 - 12 M^2 u^4 + \Lambda/2 ]}{3 u^4 [-3 q^4 u^4 + 8 M^2 u^2 - 1 ]}.
\end{equation}

Further, the condition $P(u)=0$ for the occurrence of circular orbits, gives us the energy $E$ of a particle executing circular motion from equation (\ref{35}) as
\begin{equation}\label{38}
E^2 = \frac{\epsilon [ 2f(u) - \Lambda ] [ f(u) - \Lambda ]}{18 u^4 [3 q^4 u^4 - 8 M^2 u^2 + 1 ]}.
\end{equation}
where $f(u)=3 q^4 u^6 - 12 M^2 u^4 + 3 u^2$.
\subsubsection{Motion of massive particles}

\begin{figure}[tb]
\psfrag{chi}{\scalebox{1}{$E^2$}}
\psfrag{h}{\scalebox{1}{$r/M$}}
\includegraphics[scale=.30]{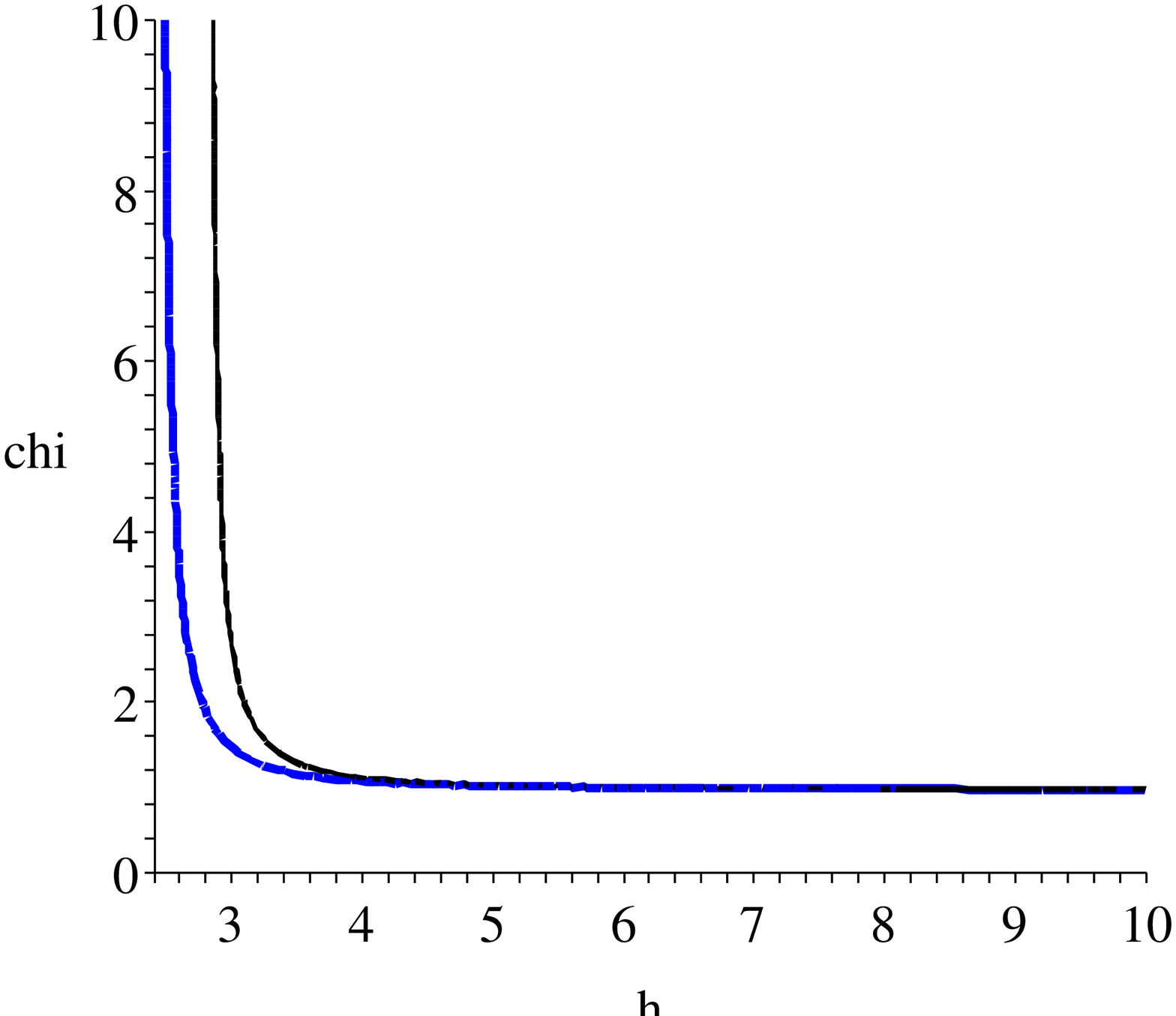}
\psfrag{Sigma}{\scalebox{1}{$\frac{L^2}{M^2}$}}
\psfrag{h}{\scalebox{1}{$r/M$}}
\includegraphics[scale=.30]{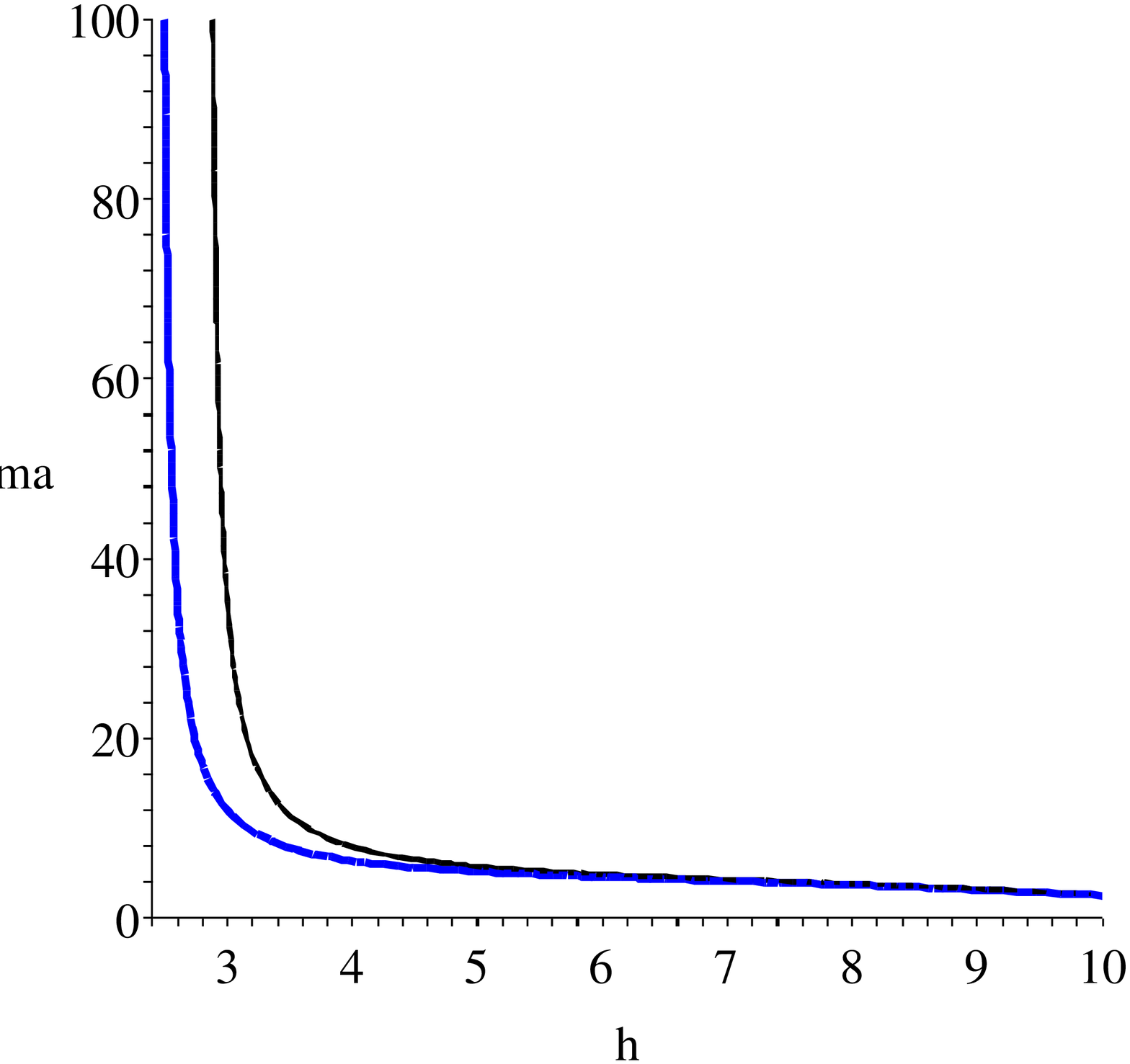}
\caption{Diagram showing the radial dependence of $E^2$ and $L^2/M^2$ for particles moving in circular orbits on the equatorial plane with $q/M = 0.5$ (black line) and $q/M = 1.414$ (blue line)}
\label{engynangmom}
\end{figure}

For massive particles, (\ref{37}) becomes
\begin{equation}\label{39}
L^2 = \frac{[ 6 q^4 u^6 - 12 M^2 u^4 + \Lambda/2 ]}{3 u^4 [-3 q^4 u^4 + 8 M^2 u^2 - 1 ]}.
\end{equation}
Similarly, we can obtain the energy of massive particles moving in circular orbits as
\begin{equation}\label{40}
E^2 = \frac{ [ 2f(u) - \Lambda ] [ f(u) - \Lambda ]}{18 u^4 [3 q^4 u^4 - 8 M^2 u^2 + 1 ]}.
\end{equation}
To calculate the radius of bound circular orbits, we need to solve a polynomial of degree $8$ in $u$, which yields eight possible values of $r/M=(1/uM)$. For an orbit of a given radius, the energy $E$ is a constant of motion. However, $E$ varies as the radius of the orbits vary. The variation of $E^2$ and $L^2/M^2$ as a function of $r/M$ for circular motion of massive particles on the equatorial plane of the black hole, is shown in Fig.~\ref{engynangmom} for different values of $q/M$.

The graphs indicate that the energy decreases as the value of $r/M$ increases for a given value of $q/M$. For a black hole with lesser value of specific charge, particles moving in orbits closer to the black hole, will have higher energy and angular momentum. However, for sufficiently large values of $r/M$, the energy and angular momentum tends to a constant value. For the case of 4-dimension, we recall that the circular orbits remain bound over a range of values of $r/M$ up to infinity, with the total energy tending to saturate to a stable magnitude as $r/M \rightarrow \infty$ \citep{Hobson}. Thus the behavior is similar even in 5-dimensions.

\subsubsection{Motion of photons}

The energy and angular momentum of the photons cannot be determined from equations (\ref{37}) and (\ref{38}), since $\epsilon=0$ for photons. Substituting $\epsilon=0$ in (\ref{22}) we obtain
\begin{equation}\label{41}
\left( \frac{du}{d\psi} \right)^2 = - q^4 u^6 + 4 M^2 u^4 - u^2 + \left( \frac{E^2}{L^2} + \frac{\Lambda}{6} \right) = Q(u).
\end{equation}
Therefore,
\begin{equation}\label{42}
Q^{\prime}(u) = - 6q^4 u^5 + 16 M^2 u^3 - 2 u.
\end{equation}
For $Q^{\prime}(u) = 0$, we have
\begin{equation}\label{43}
u^2 = \frac{4 M^2 \pm \sqrt{16 M^4 - 3 q^4}}{3 q^4}
\end{equation}
and from $Q(u) =0$ we obtain
\begin{equation}\label{44}
\frac{E^2}{L^2} = q^4 u^6 - 4 M^2 u^4 + u^2 - \frac{\Lambda}{6}.
\end{equation}
Equation (\ref{43}) indicates that the photons can have circular trajectories for only two specific values of $r/M$ for a black hole of a given mass and charge. This feature is an important distinction between the circular motion of photons and those of massive particles discussed above.

\subsection{Stability of orbits of massive particles}

\begin{figure}[tb]
\psfrag{V_eff}{\scalebox{1}{$V_eff$}}
\psfrag{h}{\scalebox{1}{$r/M$}}
\includegraphics[scale=.30]{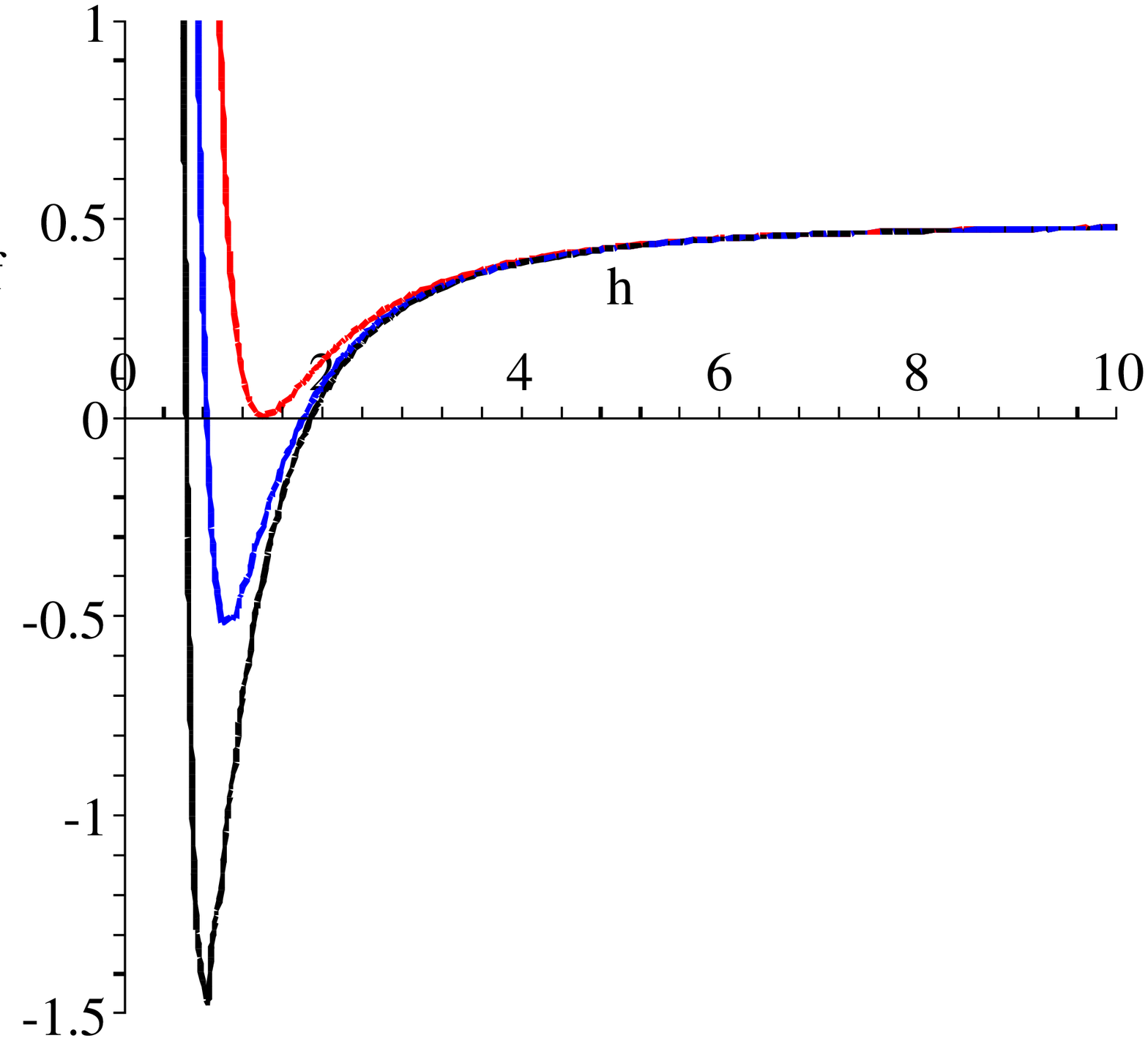}
\includegraphics[scale=.30]{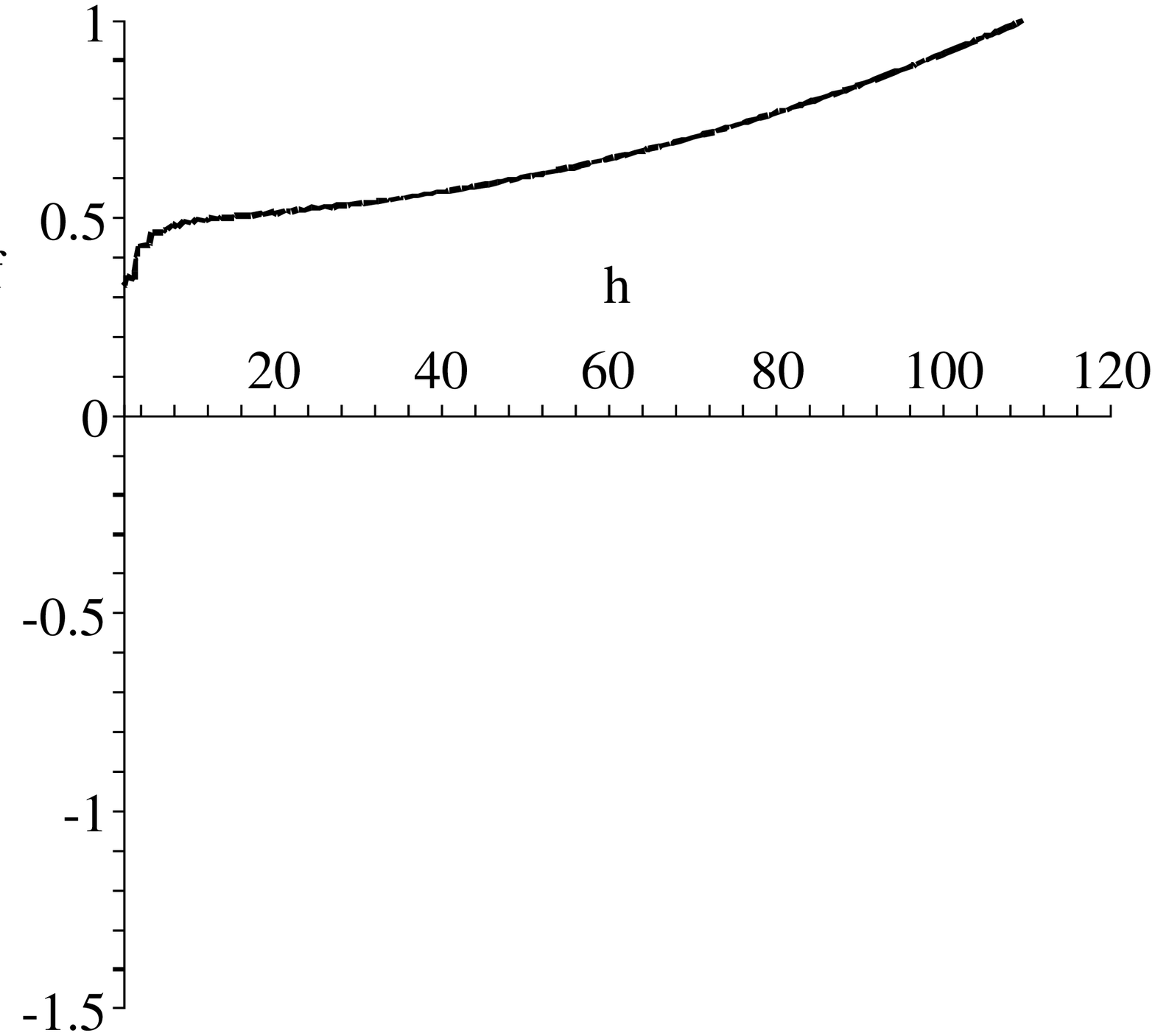}
\caption{Diagram showing the variation of the effective potential $V_{eff}$ with a variation of the specific radius of circular trajectories of massive particles for increasing values of q/M with $q/M=1.1$ (black), $q/M=1.23$ (blue) and $q/M=1.414$ (red)}
\label{stateA}
\end{figure}

The stable circular orbits occur for those values of $r$ which are located at the local minimum of the potential. The local maxima of these curves correspond to the radii of unstable circular orbits. Fig.~\ref{stateA} shows the plot of the effective potential vs $r/M$ for different values of $q/M$.

\begin{figure}[tb]
\psfrag{V_eff}{\scalebox{1}{$V_eff$}}
\psfrag{h}{\scalebox{1}{$r/M$}}
\includegraphics[scale=.30]{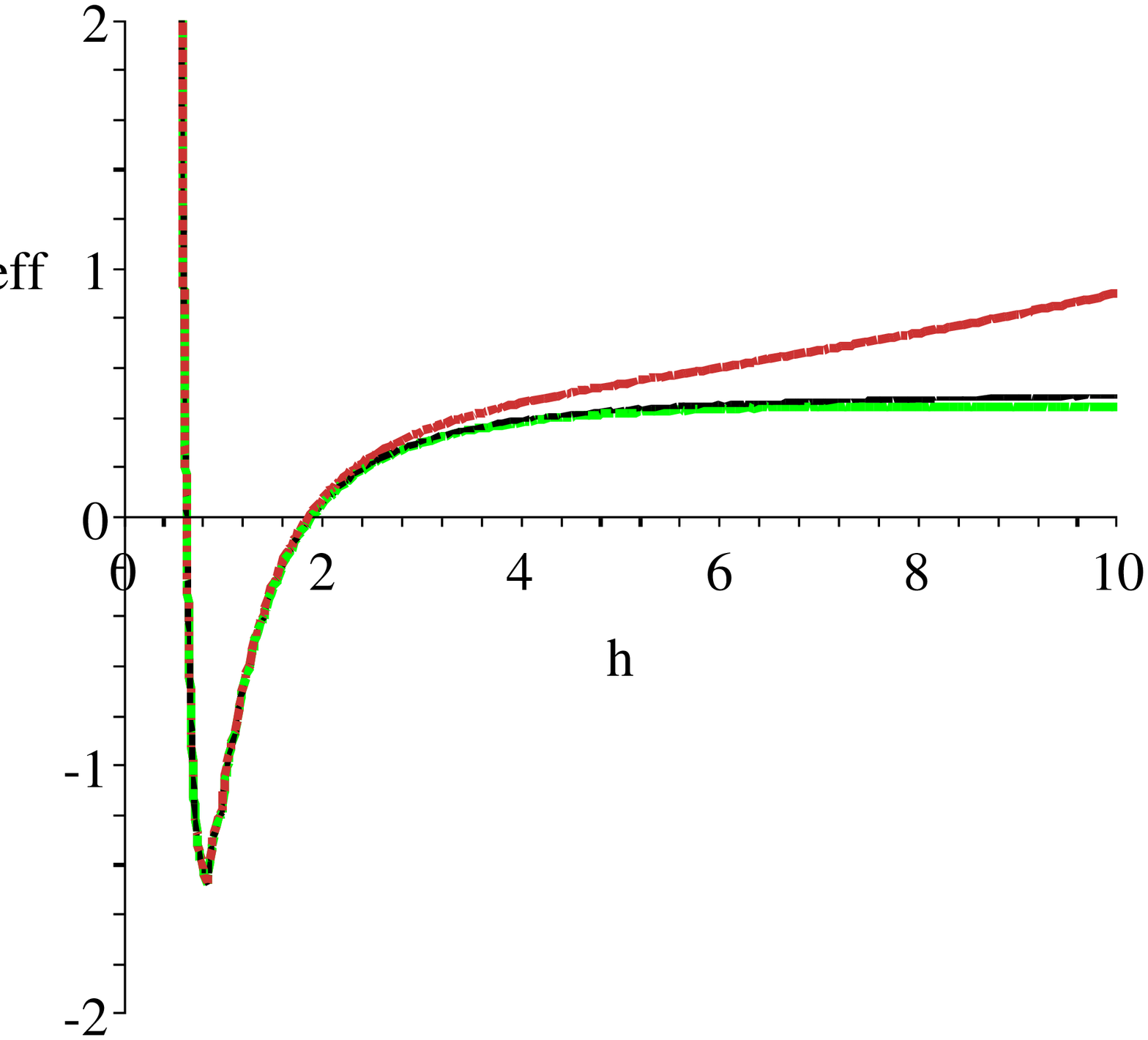}
\includegraphics[scale=.30]{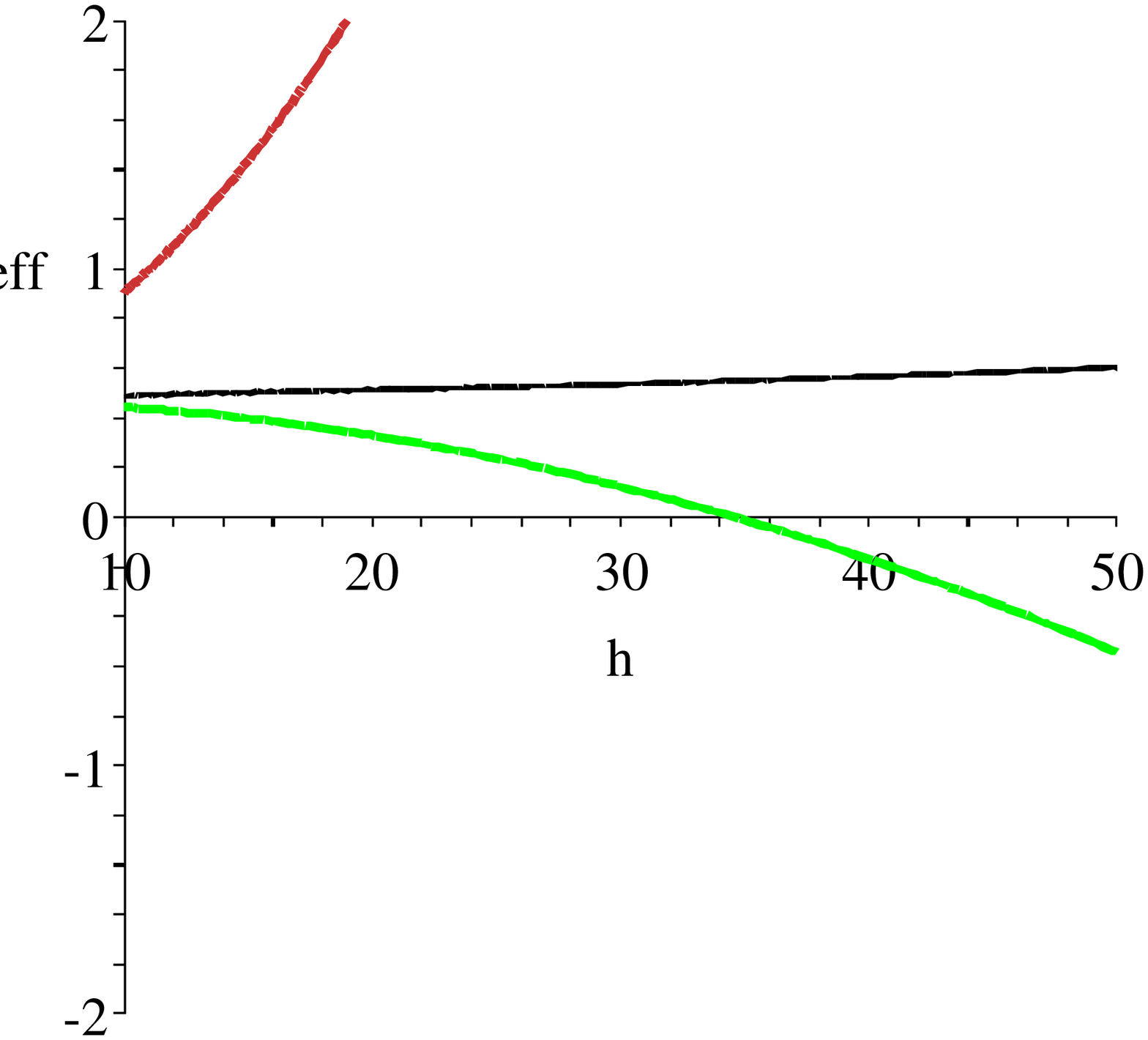}
\caption{Diagram showing the plot of effective potential $V_{eff}$ over different range of values of $r/M$ for sample values of $\Lambda=-0.05$ (orange), $\Lambda=0.-0005$ (black) and $\Lambda=0.005$ (green) in the case of circular trajectories of massive particles having $q/M = 1.1$}
\label{stateB}
\end{figure}

Fig.~\ref{stateB} shows the variation of the effective potential for different values of $\Lambda$ with $q/M = 1.1$.

The minimum radius for stable circular orbits will occur at the point of inflexion of the function $P(u)$, for which we must have $P^{\prime \prime}(u) = 0$, along with $P(u)=0$ and $P^{\prime}(u)=0$. From (\ref{36}), we get
\begin{eqnarray*}
P^{\prime\prime}(u) = - 30 q^4 u^4 + \left( 48 M^2 - \frac{12 q^4}{L^2} \right)u^2
\end{eqnarray*}
\begin{equation}\label{47}
+ \left(\frac{8 M^2}{L^2} - 2 \right) + \frac{\Lambda}{L^2 u^4}.
\end{equation}
Thus to find the radius of the innermost stable circular orbit of massive particles, we have to solve a polynomial of degree eight in $u$. However, we can reduce it to a polynomial of degree 4 with the help of the substitution $u = \sqrt{x}$. This leads us to the equation
\begin{equation}\label{47a}
30 L^2 q^4 x^4 - ( 48 M^2 L^2 - 12 q^4)x^3 - (8 M^2 - 2 L^2)x^2 - \Lambda = 0.
\end{equation}
The radius of the innermost stable circular orbit of massive particles in a five-dimensional RN-AdS$_{5}$ can be calculated from (\ref{47a}), provided $4M^2L^2 \neq q^4$ and $4M^2 \neq L^2$. These tracks will be visible to an observer only if this minimum radius is greater than the radius of the event horizon.

\subsubsection{Special case}

For $4M^2L^2 = q^4$, $4M^2 \neq L^2$ since $\Lambda < 0$, we find that
\begin{equation}\label{47d}
r_{ISCOM}^2 = \frac{\sqrt{30}Lq^2}{(4M^2 - L^2) + \sqrt{(4M^2 - L^2)^2 + 30L^2q^4 \Lambda}},
\end{equation}
indicating that in this case there will be only two possible real values of the radius, provided $(4M^2 - L^2)^2 > |30L^2q^4 \Lambda|$.

\subsection{Stability of orbits of photons}

\begin{figure}[tb]
\psfrag{V_}{\scalebox{1}{$\bar{V}_{eff}$}}
\psfrag{h}{\scalebox{1}{$r/M$}}
\includegraphics[scale=.30]{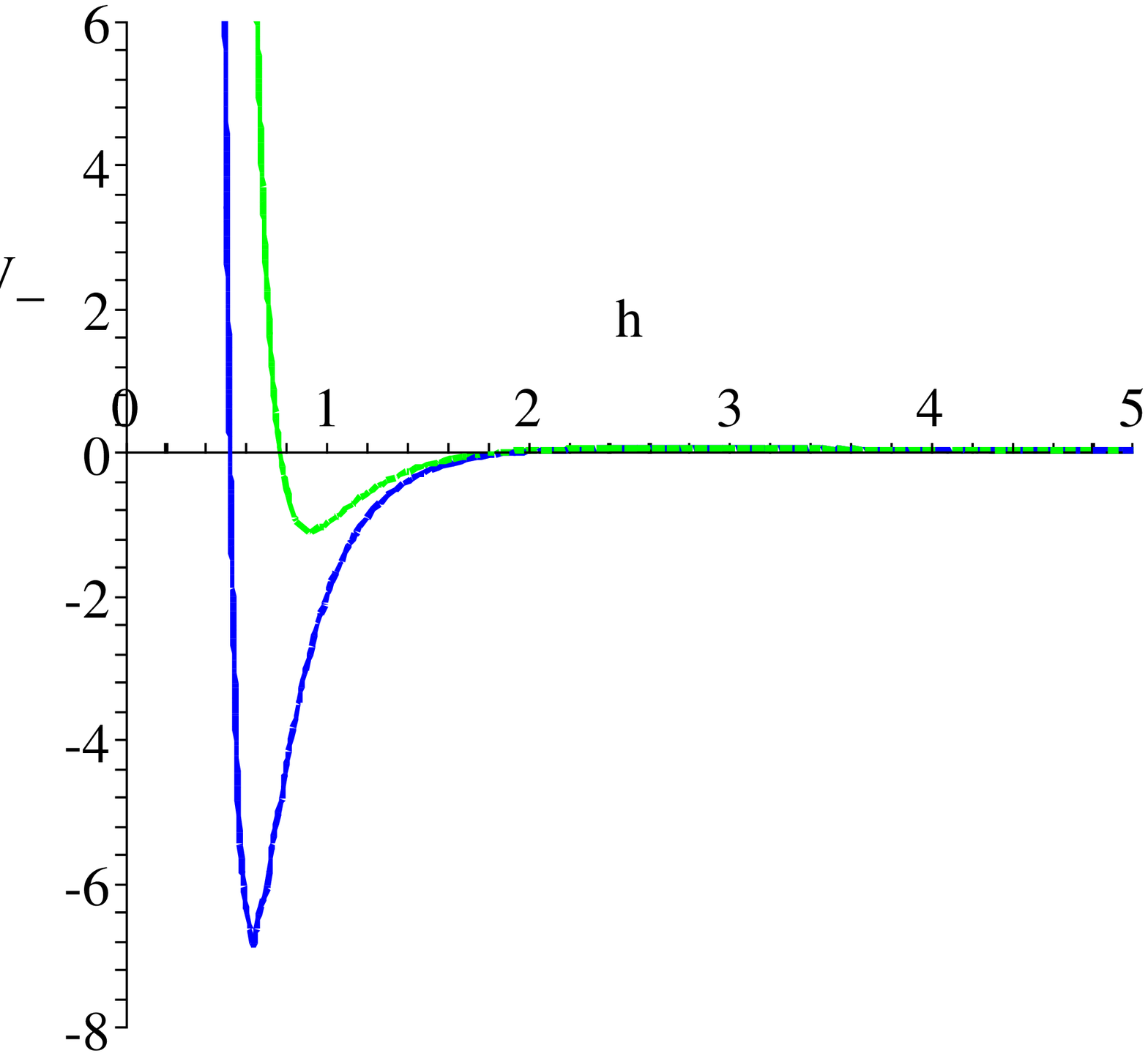}
\includegraphics[scale=.30]{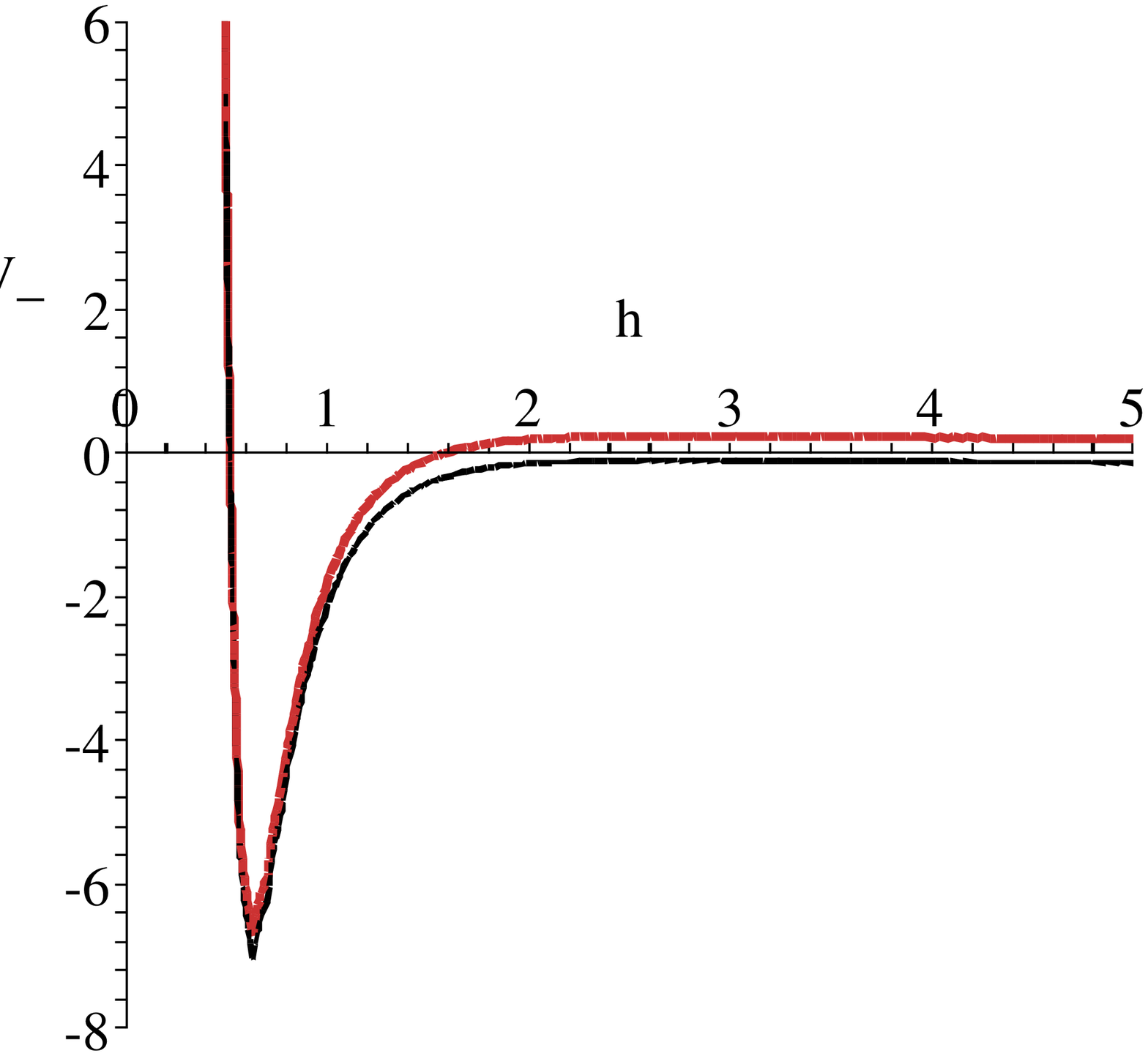}
\caption{Variation of the effective potential $\bar{V}_{eff}$ with the variation of the specific radius of circular trajectories after the transformation from $\lambda$ to $L \lambda$. On the left, $\Lambda=-0.0005$ with $q/M=1$ (blue) and $q/M=1.19$ (green). On the right, $q/M=1$ with $\Lambda=-1.0$ (orange) and $\Lambda=+1.0$ (black)}
\label{stateC}
\end{figure}

To analyze the stability of photon trajectories, we consider the first integral of the geodesic equations
\begin{equation}\label{48}
\left (\frac{dr}{d \lambda} \right)^2 + f(r)\frac{L^2}{r^2} = E^{2},
\end{equation}
which can be recasted as
\begin{equation}\label{48a}
\frac{1}{L^2}\left (\frac{dr}{d \lambda} \right)^2 + \bar{V}_{eff}(r) = \bar{E}_{eff},
\end{equation}
with
\begin{eqnarray*}
  \bar{V}_{eff} = \frac{1}{r^2} \left[ 1 - \frac{4 M^2}{r^2} + \frac{q^4}{r^4} - \frac{\Lambda r^2}{6} \right]
\end{eqnarray*}
and
\begin{eqnarray*}
\bar{E}_{eff} = \frac{E^2}{L^2}.
\end{eqnarray*}
Making the transformation $\lambda \longrightarrow L \lambda$ we can rewrite this equation in the form of an equation for effective potential $\bar{V}_{eff}$. Fig.~\ref{stateC} shows the plot of $\bar{V}_{eff}$ vs $r/M$.

When the photons describe circular trajectories, the minimum radius $r_{mc}$ of the stable circular orbits can be obtained from the condition $Q^{\prime \prime}(u) = 0$, where
\begin{equation}\label{49}
Q^{\prime \prime}(u) = - 30 q^4 u^4 + 48 M^2 u^2 - 2.
\end{equation}
Thus we get
\begin{eqnarray*}
  r_{mc} > \frac{\sqrt{15}q^2}{\sqrt{12M^2 - \sqrt{ 144M^4 - 15q^4 } }}.
\end{eqnarray*}

\section{Analysis of particle trajectories using the dynamical systems approach}
\label{s:4}

In this section, we consider the geodesic equations as a dynamical system and proceed to determine its phase space. A number of authors have used this method in various context \citep{dynsys1,dynsys2,WE}. For the analysis of the phase trajectories and the nature of critical points, we use the method adopted earlier \citep{Guha}.

Let us define the dynamical system in terms of three new variables as follows:
\begin{equation}
U=\frac{dt}{d\lambda}, \label{e9}
\end{equation}
\begin{equation}
V=\frac{dr}{d\lambda}, \label{e10}
\end{equation}
and
\begin{equation}
W=\frac{d\psi}{d\lambda}. \label{e11}
\end{equation}
The geodesics equations (\ref{e6}) to (\ref{e8}), when written in terms of these variables, reduce to the following form:
\begin{equation}\label{e12}
\frac{dU}{d\lambda} + \frac{B(r)}{A(r)}UV = 0,
\end{equation}
\begin{equation}\label{e13}
\frac{dV}{d\lambda} + A(r)B(r)U^2 - \frac{B(r)}{A(r)}V^2 + rA(r)W^2 = 0,
\end{equation}
and
\begin{equation}\label{e14}
\frac{dW}{d\lambda} + \frac{1}{r}VW = 0.
\end{equation}
These three variables are related through the first integral of geodesic equations i.e.
\begin{equation}\label{e15}
A(r)U^2 - \frac{1}{A(r)}V^2 + r^2W^2 = -\epsilon.
\end{equation}
It is not necessary to study the complete system (\ref{e12}) to (\ref{e14}), since $W$ can be determined algebraically from (\ref{e15}). Substituting for $W$ from (\ref{e15}) in (\ref{e13}), we obtain
\begin{eqnarray*}
\frac{dV}{d\lambda} + \frac{A(r)}{r}(rB(r)-A(r))U^2 + \frac{1}{rA(r)}(A(r)
\end{eqnarray*}
\begin{equation}\label{e16}
-rB(r))V^2 - \frac{\epsilon A(r)}{r} = 0.
\end{equation}
The description of the phase trajectories now reduces to the analysis of the real non-linear dynamical system defined by the equations
\begin{eqnarray}\label{e17}
\nonumber \frac{dr}{d\lambda} &=& V, \\
\frac{dU}{d\lambda} &=& H(U,V,r), \\
\nonumber \frac{dV}{d\lambda} &=& J(U,V,r),
\end{eqnarray}
where
\begin{equation}
\nonumber H(U,V,r) = - \frac{B(r)}{A(r)}U V
\end{equation}
and
\begin{eqnarray*}
\nonumber J(U,V,r) = \frac{\epsilon A(r)}{r}  - \frac{A(r)}{r}(rB(r)-A(r))U^2 \\ - \frac{1}{rA(r)}(A(r)-rB(r))V^2
\end{eqnarray*}
with $\epsilon = 1 , 0 $ for timelike and null geodesics respectively. Both $H$ and $J$ have continuous first partial derivatives for all $(U,V)$.

\subsection{Determination of the fixed points}

The fixed points of the above dynamical system are the solutions of the system (\ref{e17}), for which $V = 0$, $H = 0$ and $J = 0$. These solutions represent the equilibrium positions for the particle trajectories, which can be either stable or unstable.

Let $(U_{0},V_{0}=0,r_{0})$ represent the fixed point of the phase trajectories. We can see that $V_{0}=0$ is sufficient to satisfy $H=0$. It also implies that $r$ is constant for the fixed point. To have $J=0$ in this condition, we further need $A(r)=-f(r)=0$, for both timelike and null geodesics, where $f(r)$ is given by equation (\ref{05}). In fact, the value of $r_{0}$ can be determined from the condition $A(r)=0$, which means that $r_{0}$ is constant for a black hole of a given mass $M$ and charge $q$, when $\Lambda$ remains constant. Consequently, for a given black hole, the fixed points will lie on a definite phase plane corresponding to a constant value of $r_{0}$. Finally, the coordinate $U_{0}$ can be evaluated from
\begin{equation}\label{e18}
\frac{\epsilon A(r_{0})}{r_{0}} - \frac{A(r_{0})}{r_{0}}(r_{0}B(r_{0})-A(r_{0}))U_{0}^2 = 0,
\end{equation}
where $\epsilon = 0,1$ respectively for null geodesics and timelike geodesics. Thus we have
\begin{equation}\label{e19}
U_{0}^2=\frac{\epsilon}{(r_{0}B(r_{0})-A(r_{0}))}.
\end{equation}

For \emph{null geodesics} $\epsilon = 0$, so that
\begin{equation}\label{e20}
U_{0} = 0.
\end{equation}
Hence, for a black hole with a given $M$ and $q$, (0,0) is the fixed point on the ($U$-$V$) phase plane defined by $r_{0}= \alpha$ where $\alpha$ is a constant.

For \emph{timelike geodesics}, $\epsilon = 1$ and
\begin{equation}\label{e21}
U_{0}=\pm\sqrt{\frac{1}{(r_{0}B(r_{0})-A(r_{0}))}}.
\end{equation}
Evidently, the timelike geodesics will possess definite fixed points for a black hole of a given $M$ and $q$.

\subsection{Analysis of phase trajectories}

The phase evolution of the system in the $(U,V)$ phase plane is given by the solution of the set (\ref{e17}). We note that the equation
\begin{equation}\label{e22}
\frac{dV}{dU} = \frac{J}{H}
\end{equation}
specifies the phase evolution of the system in the $(U,V)$ phase plane, provided $H \neq 0$ at this point, which is true except at the critical point (0,0) for the null geodesics. Substituting the expressions for $H$ and $J$ in (\ref{e22}) and integrating, we obtain
\begin{equation}\label{e23}
\frac{A(r) C(r) U^3}{3 r} - \frac{\epsilon A(r) U}{r} - \frac{C(r) U V^2}{r A(r)} - \frac{B(r) U V^2}{2 A(r)} + D = 0,
\end{equation}
where $D$ is a constant of integration and $C(r) = rB(r)-A(r) $. We can rescale our coordinates so as to reduce this constant $D$ to zero and simplify to arrive at the result
\begin{equation}\label{e24}
f_{1} U^2 - f_{2} V^2 - f_{3} = 0,
\end{equation}
where $f_{1}$, $f_{2}$ and $f_{3}$ are functions of $r$. For null geodesics, we have $f_{3} = 0$.

\begin{figure}[tb]
\includegraphics[scale=.30]{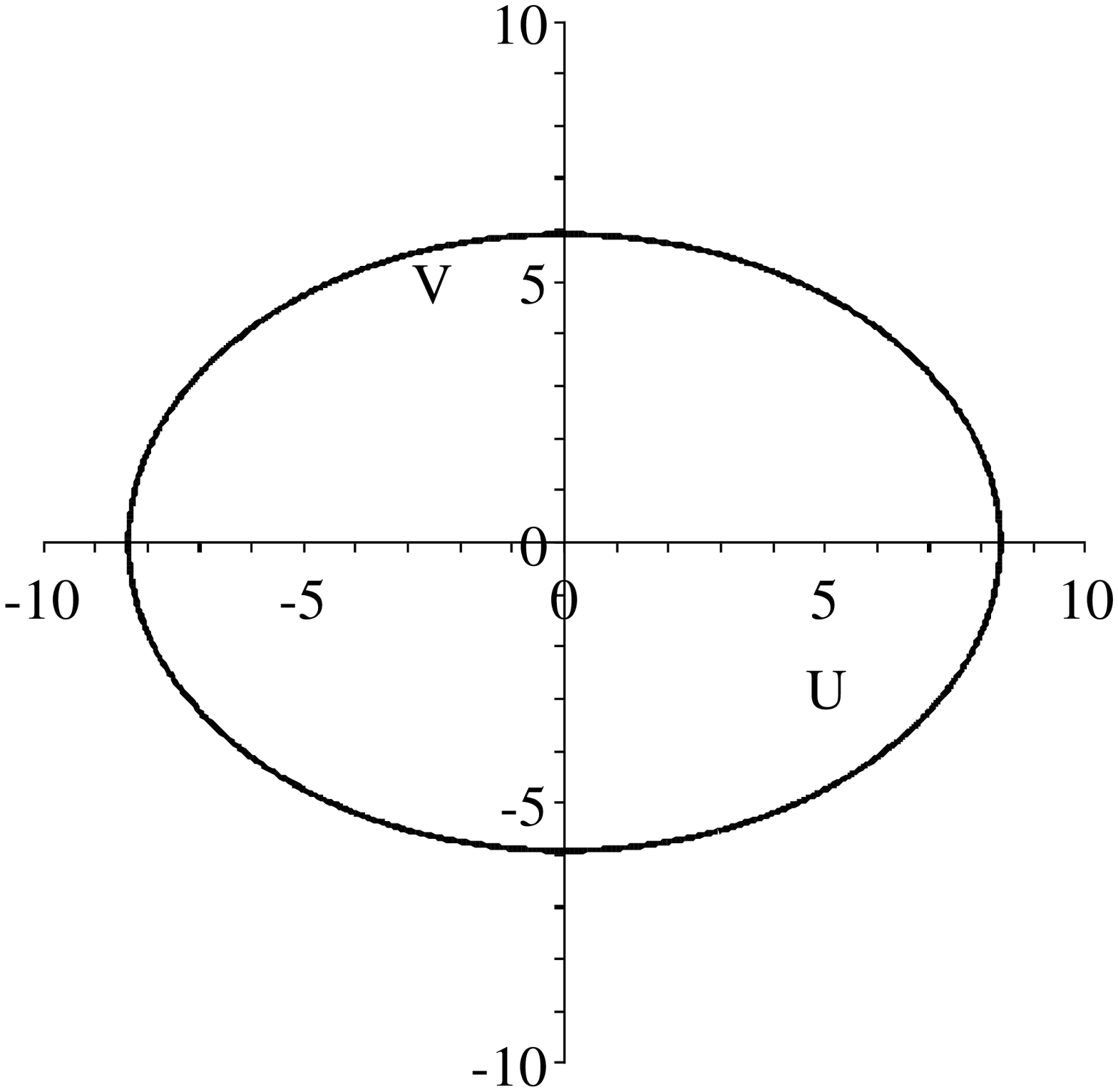}
\caption{Diagram showing the plot of sample equation $10U^2+20V^2-700=0$ for timelike geodesics}
\label{state1}
\end{figure}

\begin{figure}[tb]
\includegraphics[scale=.30]{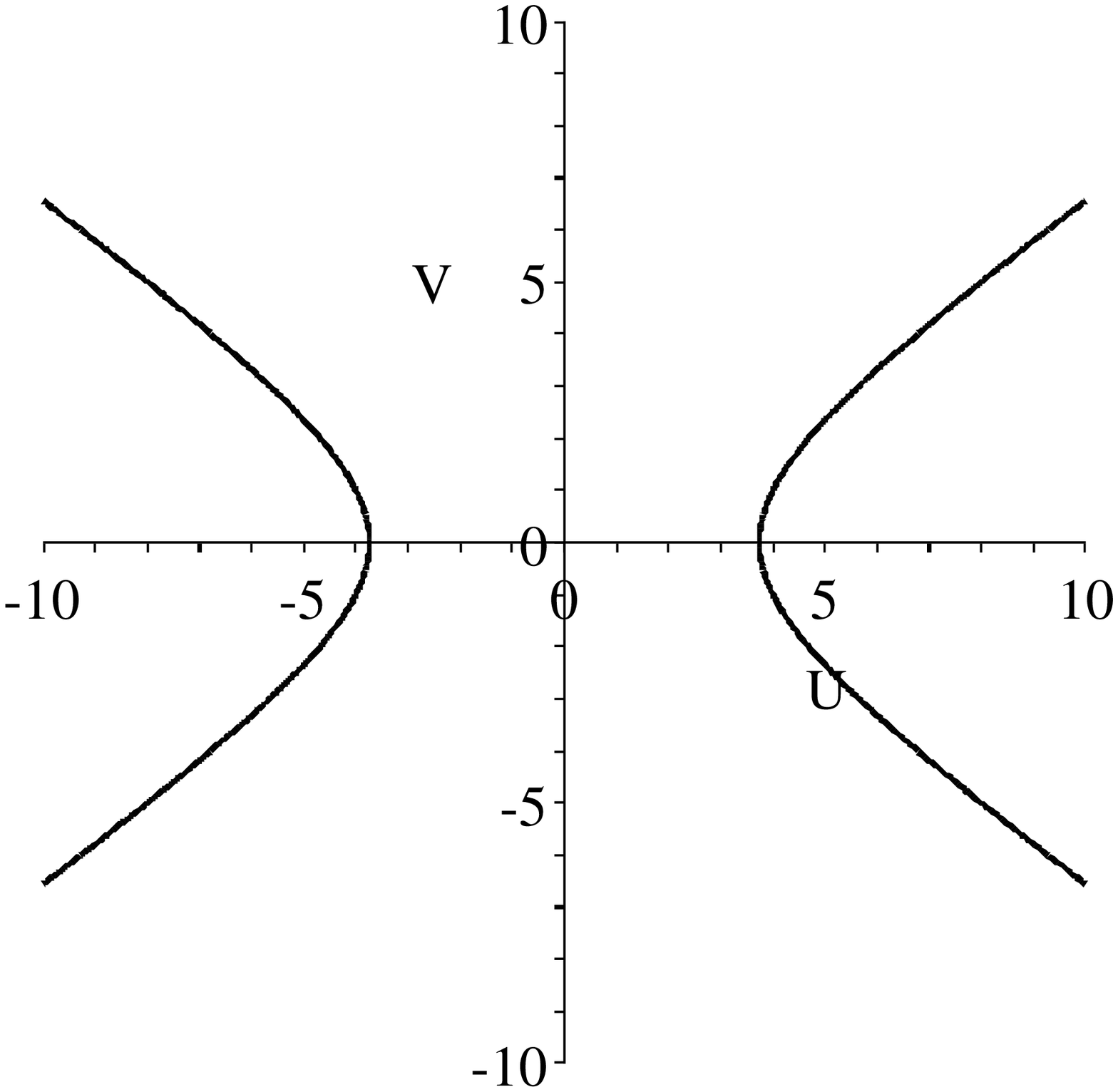}
\caption{Diagram showing the plot of sample equation $5U^2-10V^2-70=0$ for timelike geodesics}
\label{state2}
\end{figure}

Below Fig.~\ref{state1} and Fig.~\ref{state2} show possible trajectories of massive particles for different choices of parameters $f_{1}$, $f_{2}$ and $f_{3}$. We find that the trajectories are either elliptic or a pair of hyperbolas.

\begin{figure}[tb]
\includegraphics[scale=.30]{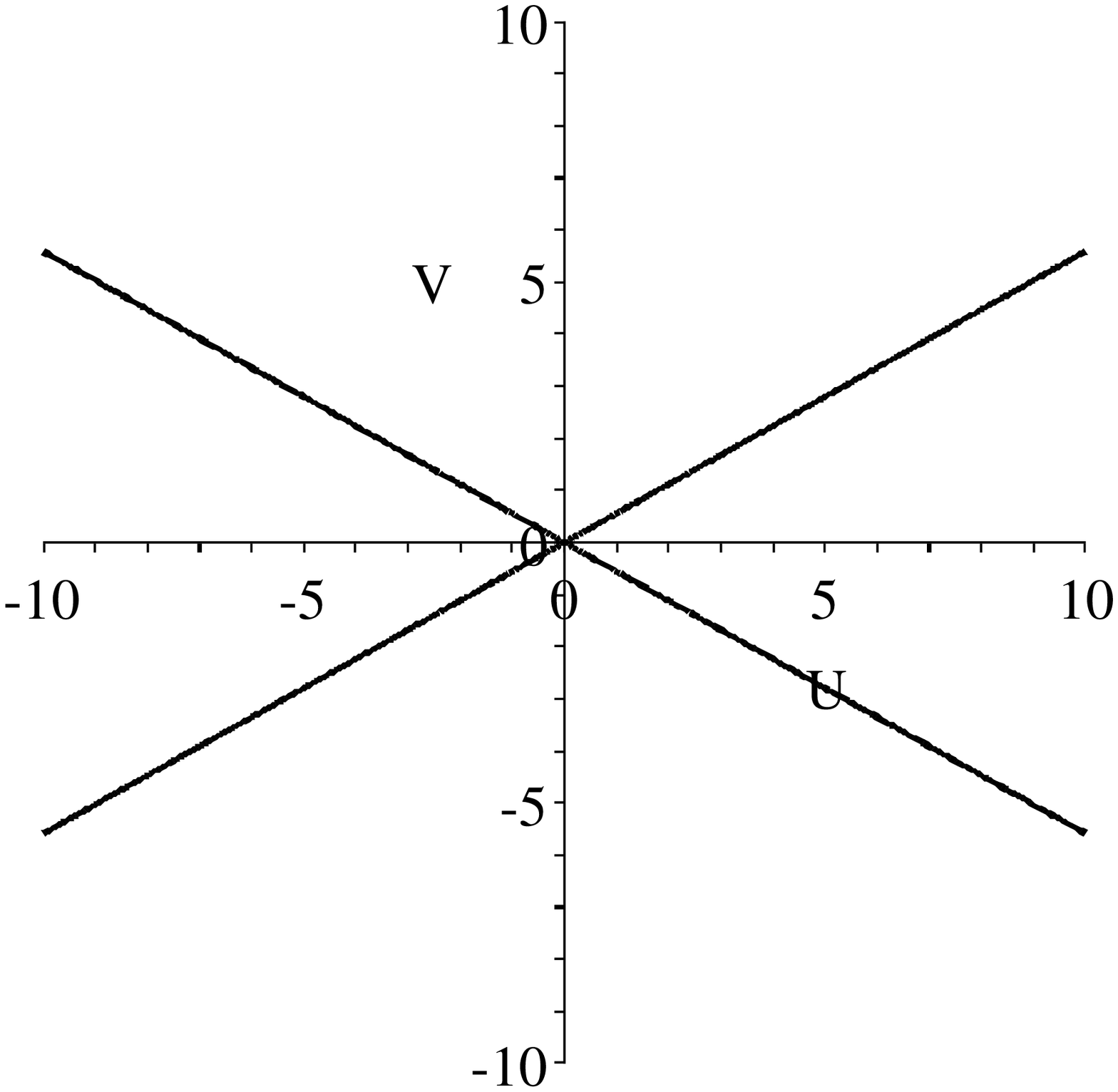}
\caption{Diagram showing the plot of sample equation $-5U^2+16V^2=0$ for null geodesics.}
\label{state3}
\end{figure}

The phase trajectories of the photons are a pair of straight lines intersecting at $(0,0)$, with their slopes changing according to the choice of the parameters $f_{1}$ and $f_{2}$. Fig.~\ref{state3} shows a sample plot of the null geodesics.

\subsection{Nature of orbits}
Our investigations using the method of dynamical systems, indicate that the geodesics of massive particles near a non-rotating charged black hole in a five-dimensional spacetime possesses definite fixed points for a given black hole and are either periodic bound or escape orbits. Moreover, for a black hole of a given charge and mass, the null geodesics possesses a unique fixed point ($U_{0}=0$, $V_{0}=0$, $r_{0}$), which correspond to the condition $\frac{dt}{d\lambda}=0$ and $\frac{dr}{d\lambda}=0$ along the geodesic curves. Hence the null geodesics are terminating orbits.

\section{Summary and conclusions}
\label{s:5}

We have studied the timelike and null geodesics of neutral particles in the field of a non-rotating, charged black hole in a five-dimensional spacetime from the point of view of effective potential formalism and the dynamical systems approach. The black hole horizons are located with the help of the plot of the effective potential. In addition to radial trajectories, we have investigated the circular trajectories for photons and massive particles, as well as the fixed points of particle trajectories. It is found that photons will trace out circular trajectories for only two distinct values of specific radius of the orbits. The radius of the innermost stable circular orbit of massive particles is totally defined in terms of their angular momentum and the mass and charge of the black hole. To determine the nature of trajectories and the fixed points, we have used the dynamical systems analysis, defining the dynamical variables in terms of the derivatives of the coordinates with respect to the affine parameter along the geodesic curves. Consequently, we have found that the geodesics of massive particles near a non-rotating charged black hole in 5D are either periodic bound or escape orbits and there are definite fixed point for these trajectories. Moreover, the null geodesics have a unique fixed point ($U_{0}=0$, $V_{0}=0$, $r_{0}$,) and these orbits are terminating orbits.

We conclude with the note that here we assumed the singularity to be a black hole and not a naked singularity. We know that black holes and naked singularities can be observationally differentiated through their gravitational lensing features. Whether these singularities can be differentiated in terms of the nature of the particle trajectories is the subject of our investigation and will be reported in a future work.

\section*{Acknowledgments}
A portion of this work was done in IUCAA, India under the associateship programme. SG and SC gratefully acknowledge the warm hospitality and the facilities of work at IUCAA.

%
%

%


%


%

\bigskip
\begin{thebibliography}{00}
\bibitem[Emparan \& Reall (2008)]{ER} Emparan, R. \&  Reall, H. S. 2008 \emph{Liv. Rev. Rel.} 2008-6 (http://www.livingreviews.org/lrr-2008-6).
\bibitem[Kanti (2004)]{Kanti} Kanti, P. 2004 \emph{Int. J. Mod. Phys. A} \textbf{19} 4899
\bibitem[Argyres et al. (1998)]{collbh} Argyres, P. C.,  Dimopoulos, S. \& March-Russell, J. 1998 \emph{Phys. Lett. B} \textbf{441} 96 \\
 Emparan, R., Horowitz, G. T. \& Myers, R. C. 2000 \emph{Phys. Rev. Lett.} \textbf{85} 499 \\
  Dimopoulos, S. \& Landsberg, G. 2001 \emph{Phys. Rev. Lett.} \textbf{87} 161602 \\
   Giddings, S. B. \& Thomas, S. 2002 \emph{Phys. Rev. D} \textbf{65} 056010 \\
    Feng, J. L. \& Shapere, A. D. 2001 \emph{Phys. Rev. Lett.} \textbf{88} 021303 \\
     Eardley, D. M. \& Giddings, S. B. 2002 \emph{Phys. Rev. D.} \textbf{66} 044011
\bibitem[Gibbons et al. (2004)]{Gibbons} Gibbons, G. W., Lu, H., Page, D. N. \& Pope, C. N. 2004 \emph{Phys. Rev. Lett.} \textbf{93} 171102
\bibitem[Sen (2005)]{Sen} Sen, A. 2005 \emph{J. High Energy Phys.} 07(2005)073
\bibitem[Frolov et al. (2003)]{Frolov} To see a comprehensive list of these works see Frolov, V. \& Stojkovic, D. 2003 \emph{Phys. Rev. D} \textbf{68} 064011 and references therein.
\bibitem[Dadhich et al. (2000)]{Dadhich} Dadhich, N. K., Maartens, R., Papodopoulos, P. \& Rezania, V. 2000 \emph{Phys. Lett. B} \textbf{487} 1
\bibitem[Maartens (2004)]{lrr} Maartens, R. 2004 \emph{Liv. Rev. Rel.} 2004-7 \\ (http://www.livingreviews.org/lrr-2004-7)
\bibitem[Page et al. (2007)]{Page} Page, D. N., Kubiznak, D., Vasudevan, M. \& Krtous, P. 2007 \emph{Phys. Rev. Lett.} \textbf{98} 061102
\bibitem[Cardoso et al. (2006)]{Cardoso} Cardoso, V., Cavaglia, M. \& Gualtieri, L. 2006 \emph{Phys. Rev. Lett.} \textbf{96} 071301
\bibitem[Konoplya et al. (2008)]{Konoplya} Konoplya, R. A. \& Zhidenko, A. 2008 \emph{Phys. Rev. D} \textbf{78} 104017
\bibitem[Cruz et al. (2005)]{Cruz} Cruz, N., Olivares, M. \& Villanueva, J. R. 2005 \emph{Class. Quant. Grav.} \textbf{22} 1167
\bibitem[Abdujabbarov et al. (2010)]{Abd} Abdujabbarov, A. \& Ahmedov, B. 2010 \emph{Phys. Rev. D} \textbf{81} 044022
\bibitem[Hioki et al. (2008)]{Hioki} Hioki, K. \& Miyamoto, U. 2008 \emph{Phys. Rev. D} \textbf{78} 044007
\bibitem[Stuchlik et al. (1991)]{Stuch} Stuchlik, Z. \& Calvani, M. 1991 \emph{Gen. Relativ. Grav.} \textbf{23} 507
\bibitem[Kraniotis (2004)]{Kraniotis1} Kraniotis, G. 2004 \emph{Class. Quant. Grav.} \textbf{21} 4743 \\
 Fujita, R. \& Hikada, W. 2009 \emph{Class. Quant. Grav.} \textbf{26} 135002
\bibitem[Kraniotis et al. (2003)]{Kraniotis2} Kraniotis, G. \& Whitehouse, S. 2003 \emph{Class. Quant. Grav.} \textbf{20} 4817
\bibitem[Hackmann et al. (2008a)]{Hackmann1} Hackmann, E. \& L\"{a}mmerzahl, C. 2008 \emph{Phys. Rev. Lett.} \textbf{100} 171101
\bibitem[Hackmann et al. (2008b)]{Hackmann1a} Hackmann, E. \& L\"{a}mmerzahl, C. 2008 \emph{Phys. Rev. D} \textbf{78} 024035
\bibitem[Hackmann et al. (2008c)]{Hackmann2} Hackmann, E., Kagramanova, V., Kunz, J. \& L\"{a}mmerzahl, C. 2008 \emph{Phys. Rev. D} \textbf{78} 124018
\bibitem[Kagramanova et al. (2006)]{Kagra} Kagramanova, V., Kunz, J. \& L\"{a}mmerzahl, C. 2006 \emph{Phys. Lett. B} \textbf{634} 465\\
Hackmann, E. \& L\"{a}mmerzahl, C. 2008 \emph{Phys. Rev. D} \textbf{78} 024035 \\ Grunau, S. \& Kagramanova, V. 2011 \emph{Phys. Rev. D} \textbf{83}, 044009
\bibitem[Gibbons et al (2002)]{Gibbons2} Gibbons, G. W., Ida, D. \& Shiromizu, T. 2002 \emph{Prog. Theor. Phys. Suppl}. \textbf{148} 284
\bibitem[Virbhadra (2009)]{Virbhadra1} Virbhadra, K.S. 2009 \emph{Phys. Rev. D} \textbf{79} 083004
\bibitem[Virbhadra et al. (2002)]{Virbhadra2} Virbhadra, K.S. \& Ellis, G. F. R. 2002 \emph{Phys. Rev. D} \textbf{65} 103004
\bibitem[Iorio (2011)]{Iorio} Iorio, L. 2011 arXiv:1112.3520 [gr-qc]
\bibitem[Carroll (2004)]{Carroll} Carroll, S. 2004 {\it Spacetime and Geometry} (Addison Wesley)
\bibitem[Cardoso et al. (2009)]{Cardoso1} Cardoso, V., Lemos, M. \& Marques, M. 2009 \emph{Phys. Rev. D} \textbf{80} 127502
\bibitem[Chandrasekhar (1983)]{Chandra} Chandrasekhar, S. 1983 {\it The Mathematical Theory of Black Holes} (Oxford University Press)
\bibitem[Uzan et al. (2001)]{dynsys1} For analysis using the dynamical systems method in cosmology see for example: Uzan, J.-P. \& Lehoucq, R. 2001 \emph{Eur. J. Phys.} \textbf{22} 371
\bibitem[Dahia et al. (2007)]{dynsys2} For the study of geodesics using the dynamical systems analysis, see for example: Dahia, F., Romero, C., da Silva, L. F. P. \& Tavakol, R. 2007 \emph{J. Math. Phys.} \textbf{48} 072501 \\
     Dahia, F., Romero, C., da Silva, L. F. P. \& Tavakol, R. 2008 \emph{Gen. Rel. Grav.} \textbf{40} 1341
\bibitem[Hartle (2003)]{Hartle} Hartle, J. B. 2003 {\it Gravity, An Introduction to Einstein's General Relativity} (Pearson Education)
\bibitem[Hobson et al. (2006)]{Hobson} Hobson, M. P., Efstathiou, G. \& Lasenby, A. N. 2006 {\it General Relativity: An Introduction for Physicists} (Cambridge University Press)
\bibitem[Wainwright et al. (1997)]{WE} Wainwright, J. \& Ellis, G. F. R. 1997 \emph{Dynamical Systems in Cosmology} (Cambridge: Cambridge University Press)
\bibitem[Guha et al. (2010)]{Guha} Guha, S. \& Chakraborty, S. 2010 \emph{Gen. Relativ. Grav.} \textbf{42} 1739
\end{thebibliography}

%

\end{document}